\documentclass[10pt,journal,twocolumn]{IEEEtran}
\newif\ifCLASSOPTIONromanappendices \CLASSOPTIONromanappendicestrue
\usepackage{tikz}
\usepackage[siunitx]{circuitikz}
\usepackage[utf8]{inputenc}
\usepackage{fontenc}
\usepackage{hyperref}
\usepackage{a0size}         
\usepackage{amssymb}
\usepackage{amsmath}
\usepackage{upgreek}
\usepackage{multicol}
\usepackage[english]{babel}     
\usepackage{epsfig}
\usepackage{bm}
\usepackage{oplotsymbl}
\usepackage{bbm}
\usepackage{enumitem}
\usepackage{amsfonts,color,amsthm,amsmath, lscape,graphics}
\usepackage{subfiles}
\usepackage[all]{hypcap}
\usepackage{comment}
\usepackage{algorithm} 
\usepackage{algpseudocode}

\usepackage{graphicx}
\usepackage[labelformat=simple]{subcaption}
\usepackage{mwe}

\usepackage{caption}
\usepackage{epstopdf}
\usepackage{algorithm} 
\usepackage{algpseudocode}
\usepackage{cite}
\usepackage{dblfloatfix}

\definecolor{awesome}{rgb}{1.0, 0.13, 0.32}

\addto\captionsenglish{}

\newcommand*\xbar[1]{%
  \hbox{%
    \vbox{%
      \hrule height 0.5pt % The actual bar
      \kern0.2ex%         % Distance between bar and symbol
      \hbox{%
        \kern-0.1em%      % Shortening on the left side
        \ensuremath{#1}%
        \kern-0.1em%      % Shortening on the right side
      }%
    }%
  }%
}
\usepackage{booktabs}
\usepackage{multicol}
\usepackage{graphicx}  
\theoremstyle{plain}

\usepackage{float}

\usepackage{multirow} 
\usepackage{relsize}  
\usepackage{textcomp, mathtools}

\usepackage{booktabs}
\usepackage{multirow}
\usepackage{adjustbox}

\definecolor{github-link}{RGB}{0,0,139}

\usepackage{bm}   
\usepackage{pifont}
\usepackage{etoolbox}

\begin{document}
\title{Next-slot OFDM-CSI Prediction:\\ Multi-head Self-attention or State Space Model?}
\author{Mohamed Akrout, \IEEEmembership{Member, IEEE}, Faouzi~Bellili, \IEEEmembership{Member, IEEE},\\Amine Mezghani, \IEEEmembership{Member, IEEE}, Robert W. Heath, \IEEEmembership{Fellow, IEEE}
 %\vspace{0.1cm}
%\\\small E2-390 E.I.T.C,  75 Chancellor's Circle  Winnipeg, MB, Canada, R3T 5V6.
 % \vspace{0.1cm}
 % \\\small Emails:  tadeleb@myumanitoba.ca, Faouzi.Bellili@umanitoba.ca, and Ekram.Hossain@umanitoba.ca.
  %\vspace{0.3cm}
\thanks{The authors are with the Department of Electrical and Computer Engineering at the University of Manitoba, Winnipeg, MB, Canada (emails:akroutm@myumanitoba.ca, \{Faouzi.Bellili,Amine.Mezghani\}@umanitoba.ca). R.~W.~Heath is at the University of California, San Diego (email: rwheathjr@ucsd.edu). This work was supported by the Discovery Grants Program of the Natural Sciences and Engineering Research Council of Canada (NSERC) and the US National Science Foundation (NSF) Grant No. ECCS-1711702 and CNS-1731658.}}

\maketitle
%%%%%%%%%%%%%%%%%%%%%%%%%%%%%%%%%%%%
%% Abstract
%%%%%%%%%%%%%%%%%%%%%%%%%%%%%%%%%%%%
\begin{abstract}
The ongoing fifth-generation (5G) standardization is exploring the use of deep learning (DL) methods to enhance the new radio (NR) interface. Both in academia and industry, researchers are investigating the performance and complexity of multiple DL architecture candidates for specific one-sided and two-sided use cases such as channel state estimation (CSI) feedback, CSI prediction, beam management, and positioning. In this paper, we set focus on the CSI prediction task and study the performance and generalization of the two main DL layers that are being extensively benchmarked within the DL community, namely, multi-head self-attention (MSA) and state-space model (SSM). We train and evaluate MSA and SSM layers to predict the next slot for uplink and downlink communication scenarios over urban microcell (UMi) and urban macrocell (UMa) OFDM 5G channel models. Our numerical results demonstrate that SSMs exhibit better prediction and generalization capabilities than MSAs only for SISO cases. For MIMO scenarios, however, the MSA layer outperforms the SSM one. While both layers represent potential DL architectures for future DL-enabled 5G use cases, the overall investigation of this paper favors MSAs over SSMs.

%Further, suggesting that the attention mechanism is still are the dominant architecture for sequence modeling For MIMOand hence can be considered as a potential DL architecture for future DL-enabled 5G functions. \textcolor{red}{should be revised once all results are done}
\end{abstract}
\begin{IEEEkeywords}
 CSI prediction, OFDM, slot, 3GPP channel models, multi-head self-attention, state space models.
\end{IEEEkeywords}
% https://arxiv.org/pdf/2309.13414.pdf
%One of the key advantages of state-space models is their simple recurrence, which enables efficient acceleration. In fact, this recurrence allows for an asymptotic computational complexity of only O(T log T), which is significantly better than the O(T2) complexity of traditional full-attention approaches [8]. A natural question would be whether SSM achieves this speedup with certain sacrifices in model capacity or memory property. It is currently unclear whether the state-space model’s linear architecture with layerwise nonlinearity possesses sufficient expressive capacity to approximate any target sequence-tosequence relationship. This knowledge would be important to answer pertinent questions regarding the model’s ability to handle the complexity of real-world datasets characterized by diverse and intricate sequence relationships. In particular, considering the speed advantage of SSM over attentionbased transformers, the universal approximation property impacts whether a state-space model could be a suitable replacement for a transformer.

\section{Introduction}\label{Section 1}
\subsection{Background and motivation}
Generative artificial intelligence (GenAI) has recently emerged as a result of research advancements in deep learning (DL), with a promising potential to transform the technological future across numerous areas. Specifically, large language models (LLMs) and large multi-modal models (LMMs) developed within the field of natural language processing and computer vision research communities are driving innovation by enhancing automation, language translation services, and human-computer interaction (cf. \cite{yang2024harnessing} for a comprehensive overview). While GenAI is being progressively adopted by different industries, some research studies at the intersection of DL and wireless communication proposed the use of LLMs as part of self-organizing networks (SONs) \cite{bariah2024large}. These networks are expected to be highly autonomous and adaptive as they continuously optimize their functions and parameters depending on the communication conditions and user demands. To accommodate such high flexibility, GenAI for wireless communication comes into play as a key technology to generate personalized communication parameters according to network patterns and KPIs learned from massive Telecom datasets. Such AI generation can target estimation or prediction of parameters pertaining to either the physical or the network layer depending on the nature of the collected datasets and the considered downstream tasks at hand.

In this context, one of the key AI use cases considered in the recent 3rd Generation Partnership Project (3GPP) 17 and 18 releases is the channel state information (CSI) prediction \cite{lin2023overview}. An important issue with the current CSI reporting system in the new radio (NR) interface is the delay between the CSI's reporting time and the moment the CSI is actually used. This delay makes the CSI outdated due to the channel time variations. The rate at which the CSI loses its relevance depends on the channel properties and is amplified by the speed of user equipment. To address this challenge, both model-based and learning-based channel prediction techniques leverage historical CSI correlations to forecast future channel conditions and/or realizations. To capture the dynamic behavior of the channel, model-based methods employ linear extrapolation \cite{kim2020massive}, sum-of-sinusoids \cite{andersen1999prediction}, and autoregressive models \cite{peng2017channel} (see \cite{jiang2019neural} for a comprehensive overview). Due to their low complexity, learning-based approaches using deep neural networks (DNNs) stood out in the 3GPP's 5G standard discussions as a promising low-complexity strategy to predict the channel and mitigate the impact of outdated CSI. Indeed, when the channel blockage model is not available, model-based methods cannot accurately capture the large number of blockage possibilities. Such a setting is equivalent to having a non-stationary channel whose transitions can not be predicted well using linear methods.

%I think model based approaches, for example, are not good predictions when there are the possibilities of blockage. This is equivalent to having a non-stationary channel - the transitions can not be predicted well using linear methods.

\subsection{Related work}

Many standard architectures of DNNs have been investigated for multiple-input-multiple-output (MIMO) channel predictions. Multi-layer perception (MLP) was used in \cite{yang2020deep} to rely on the uplink CSI to predict the downlink one under the assumption of a direct user-channel matrix relationship, which is not always applicable. Convolutional neural networks (CNNs) and long short-term memory (LSTM) networks were employed in  \cite{luo2018channel}, yielding notable prediction performance compared to traditional methods like maximum likelihood and minimum mean squared error (MMSE). To improve upon this, recurrent neural networks (RNNs) were combined with CNNs for feature extraction, outperforming standalone CNNs for channel prediction \cite{wang2019ul}. Because the channel is complex-valued and CNNs are designed for real-valued processing only, a complex-valued 3D CNN was proposed for CSI prediction in \cite{zhang2020cv}, improving the CSI prediction accuracy of real-valued networks. Graph neural networks (GNNs) have also been applied to CSI prediction as a multivariate time-series forecasting problem \cite{mourya2024spectral} by exploiting the spectral
and temporal correlations of the historical CSI. To mitigate the sequential processing nature of LSTM networks, transformers rely on the attention mechanism to process entire sequences of input data in parallel, thereby significantly reducing training times and enabling the model to scale with the amount of data more effectively \cite{vaswani2017attention}. When applied to CSI prediction, transformers outperformed all other DDN architecture in terms of both mean square error and achievable rate \cite{jiang2022accurate}.

\begin{figure}[h!]
    \begin{subfigure}{.48\textwidth}
        \centering
        %\hspace{0.5cm}
        \includegraphics[scale=0.45]{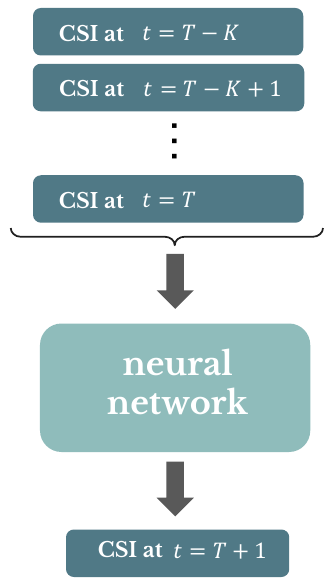}
        \caption{CSI prediction}
        \label{fig:csi-prediction}
    \end{subfigure}
    %\hspace*{3cm}
    
    \begin{subfigure}{.48\textwidth}
        \vspace*{0.5cm}
        \centering
        \includegraphics[scale=0.45]{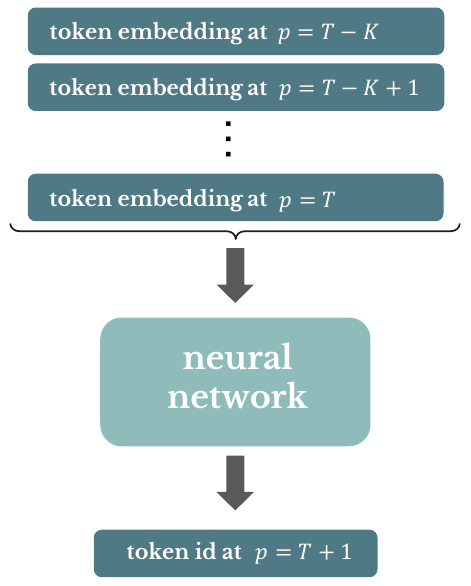}
        \caption{Next-token prediction}
        \label{fig:next-token-prediction}
    \end{subfigure}
\caption{The input-output similarity of a neural network between (a) CSI prediction and (b) next token prediction.}
    \label{fig:similarity-CSI-LLM}
    %\vspace{-0.5cm}
\end{figure}

The collaborative development of 3GPP standard involving research bodies and industry stakeholders is currently investigating the performance of multiple DNN models in terms of floating point operations per second (FLOPS) and memory complexity. Because the CSI prediction problem follows a one-sided model (i.e., inference can be conducted by one side: either the base station or the UE)\footnote{This is to be opposed to two-sided models in the 3GPP standard where the first part of the inference runs on the UE side and the second part runs on the base station side or vice-versa (e.g., encoder-decoder models).}, the performance of deployable models depends heavily on specific UE/base station vendors' hardware. For this reason, it is of utmost importance in practice to investigate the prediction capability of AI layers and avoid stacking dozens of them with the only goal of claiming state-of-the-art performance at the cost of FLOPS and memory complexity. From a wireless communication perspective (i.e., non-language application), the AI models in the 3GPP standard should benefit from the recent architectures of AI layers which have proven effective for GenAI models, namely, the \textit{multi-head self-attention} (MSA) layer \cite{vaswani2017attention} and the \textit{state space model} (SSM) layer \cite{gu2023mamba}. To see why this is possible, Fig. \ref{fig:similarity-CSI-LLM} depicts the striking similarity between CSI prediction in orthogonal frequency division multiplexing (OFDM) systems and next token prediction for LLMs. Specifically, $K$ input word embeddings obtained after tokenization and present at the position interval $p\in[T-K, T]$  are analogous to $K$ input CSI at the time interval $t\in[T-K, T]$. In other words, the equivalence becomes more apparent when the token positions are substituted by the CSI (e.g., OFDM slot) position in time.

The AI community is currently benchmarking these two layers for LLMs \cite{jelassi2024repeat}, vision-related tasks including classification of images \cite{nguyen2022s4nd} and videos \cite{islam2022long}, and graph-related tasks \cite{wang2024graph}, just to name a few. We believe it is also timely that the communication community examines how these layers can be leveraged for CSI prediction. If the vision of AI for wireless is to be core component in future-generation communication systems rather than a specific functionality among other ones, it is important to understand the capabilities of AI layers in terms of both performance and FLOPS as well as area and power consumption metrics when AI models are implemented on FPGAs or ASICs. This is particularly important to pursue because the 3GPP discussions are still ongoing and no final decision about the AI models have been made yet. % These two Specifically, two AI layers are being benchstate space model layer

%Significant research has been dedicated to enhancing the efficiency of Transformers. While Transformers are undoubtedly powerful, they require substantial resources and data. Innovations such as Flash Attention, RetNet, and several others offer promising prospects, yet the Transformer architecture continues to dominate. In this paper review, we will discuss a novel architecture named Mamba.
%In this work, we focus
%questionable performance between attention and state space models
%, namely, 

%  the building blocks of LLMs and not the LLMs themselves, unless one is interested in plug-and-play approaches. And I see semantic or intent-based communication as an excuse to blindly apply LLMs.

%, and hence propritery models for each vendor, unified model used by all vendors

%the hardware device, weak companies are against unified vendors.
%hardware? Difference Between Hardware and

%, including telecom operators, equipment manufacturers, and other entities. 
%As of my last update in April 2023, the 3GPP (3rd Generation Partnership Project) standards are ongoing and not "done" in the sense that they are continuously being developed and updated
%This ensures that the standards meet the evolving needs of mobile communication.
%As the 3GPP standard 

\subsection{Contribution}\label{subsec:contrib}

We study the prediction capabilities of MSA and SSM layers for CSI prediction at the UE side as a one-sided model defined in the 3GPP standard. Different from the aforementioned work, our goal is to neither beat other DNN architectures nor obtain state-of-the-art results by cascading DNN layers at the cost of higher FLOPS. For this reason, we set focus on shadow networks with either one single MSA or SSM layer and examine their CSI predictive ability. By doing so, we provide insights into the performance of MSA and SSM layers and their competitiveness to be considered by the industry among the deployable AI models on UE devices. Toward this goal, we first define the task of predicting the next-slot OFDM-CSI and describe the parameters of the 5G wireless channels used for training and evaluation, namely, urban microcell (UMi) and urban macrocell (UMa). We then conduct an exhaustive empirical comparison between MSA and SSM layers for both in-distribution (ID) and out-of-distribution (OOD) evaluations as a function of the SNR and speed of the UE. This is because rigorous investigations of AI models must examine the trade-off between generalization and accuracy \cite{akrout2023multilayer,akrout2023domain}. Our empirical investigation reveals the following main results:
\begin{itemize}
    \item For SISO communication scenarios: SSMs exhibit better generalization capabilities in terms of SNR and user speeds compared to MSAs. For MIMO communication scenarios, however, MSAs outperform SSMs in both ID and OOD evaluations.
    \item Diversifying communication scenarios (i.e., many SNR levels within the training dataset) over which DNNs are trained for slot prediction is only beneficial for MSAs for SISO scenarios. This diversification has a negative impact on the CSI prediction MSE for MIMO scenarios where DNNs with lower SNR levels have a lower MSE performance for CSI prediction. This can be justified by the fact that introducing noise as a data augmentation technique to the training samples prevents overfitting \cite{zhong2020random}, improves robustness \cite{lopes2019improving}, and is equivalent to Tikhonov regularization \cite{bishop1995training}. Such dataset diversification confirms the challenge of choosing the training dataset and its parameters to train DNNs. This adds on top of the data/model agreement difficulty between vendors in the context of future AI-enabled communication use cases.
    %\item This can be justified by the fact that MSAs run a pairwise comparison of each signal component in the signal sequence. This is to be opposed to SSMs whose signal sequence compression into a fixed-size memory is not equally impacted by diversified communication scenarios in the dataset.
    %\item  \textcolor{red}{add another finding for multi-user experiments}
    %\item ok \textcolor{red}{data collection challenges but also which combinasion to consider for best performance}
\end{itemize}
\noindent The code to train and test MSA and SSM layers is available at {\scriptsize{\href{https://github.com/makrout/Next-Slot-OFDM-CSI-Prediction}{\textcolor{github-link}{\texttt{https://github.com/makrout/Next-Slot-OFDM-CSI-Prediction}}}}}.

\subsection{Outline}
We structure the rest of this paper as follows. In Section~\ref{sec:preliminaries}, we introduce the relevant background of MSA and SSM layers. In Section \ref{sec:prediction}, we define the next-slot OFDM-CSI prediction task and present the parameters of the 3GPP OFDM channel models. Our simulation results are presented in Section \ref{sec:results} for both SISO and MIMO communications, from which we draw out some concluding remarks In Section \ref{sec:conclusion}. %\textcolor{red}{improve after multi-user}

\section{Background}\label{sec:preliminaries}
In this section, we review the architecture of the MSA and SSM layers at the detail needed for their comprehensive exposition and comparison.

\subsection{Multi-head attention layer}
Let $\bm{X} \in \mathbb{R}^{N\times D}$ be the input sentence, where $N$ is the sequence length and $D$ is the embedding dimension. Let also $D_h$ denote the dimension of each self-attention head (a.k.a., the query size) and $H = D/D_h$ be the number of heads. A self-attention layer starts by computing query, key, and value matrices $\bm{Q}$, $\bm{K}$ and $\bm{V}$ from $\bm{X}$ using linear transformations:
\begin{subequations}\label{eq:attention-step1}
    \begin{align}
        \bm{Q} &= \bm{X}\,\bm{W}_{{q}},\\
        \bm{K} &= \bm{X}\,\bm{W}_{{k}},\\
        \bm{V} &= \bm{X}\,\bm{W}_{{v}}.
    \end{align}
\end{subequations}
where $\bm{W}_{{q}} \in \mathbb{R}^{D\times D_h}$, $\bm{W}_{{k}} \in \mathbb{R}^{D\times D_h}$, and $\bm{W}_{{v}} \in \mathbb{R}^{D\times D_h}$ are learnable parameters. Eq. (\ref{eq:attention-step1}) can be rewritten in a compact form as
\begin{equation}
    [\bm{Q},\bm{K}, \bm{V}] = \bm{X}\,\bm{W}_{{qkv}},
\end{equation}
where $\bm{W}_{{qkv}} \in \mathbb{R}^{D\times 3\,D_h}$ is an overall learnable parameter matrix. The attention map $\bm{M} \in \mathbb{R}^{N \times N}$ is then computed by scaled inner products from $\bm{Q}$ and $\bm{K}$ and normalized by the softmax function as follows:
%try to plot the attention map for ID and OOD scenarios, follow this link https://towardsdatascience.com/deconstructing-bert-part-2-visualizing-the-inner-workings-of-attention-60a16d86b5c1
\begin{equation}\label{eq:attention-map}
    \bm{M} = \textrm{softmax}\left(\frac{\bm{Q}\,\bm{K}^\top}{\sqrt{D_h}}\right).
\end{equation}
Here, the $ij$th entry, $M_{ij}$, in $\bm{M}$ represents the attention score between $\bm{Q}_i$ and $\bm{K}_j$. The self-attention operation is then applied on the value vectors to produce the output matrix
\begin{equation}\label{eq:self-attention-operation}
    \bm{O} = \bm{M}\,\bm{V}~\in\mathbb{R}^{N \times D_h}.
\end{equation}
Finally, the output $\bm{Y} \in \mathbb{R}^{N \times D}$ of the self-attention layer is calculated by a learnable linear
projection $\bm{W}_{\textrm{proj}} \in \mathbb{R}^{D \times D}$ for the concatenated self-attention outputs of each head, i.e.:
\begin{equation}\label{eq:linear-projection-map-output}
    \bm{Y} = [\bm{O}_1, \bm{O}_2, \dots, \bm{O}_H]\,\bm{W}_{\textrm{proj}}.
\end{equation}
Overall, the MSA layer can be seen as a learnable module that takes an input $\bm{X}$ and returns an output $\bm{Y}$ of the same dimension. Note, however, that both $\bm{X}$ and $\bm{Y}$ are divided logically among $H$ heads. Consequently, different segments of $\bm{X}$ and $\bm{Y}$ are able to learn the correlation patterns of some input chunks in relation to the other ones within the sequence. This division enables the multi-head attention layer to acquire richer correlation patterns within the input sequence $\bm{Y}$.

\noindent In terms of computational complexity, the FLOPS of the MSA layer is divided across four steps:
\begin{itemize}
    \item[$i)$] the three linear projections in (\ref{eq:attention-step1}) with complexity $3\,N\,D^2$,
    \item[$ii)$] the computation of the attention map $\bm{M}$ in (\ref{eq:attention-map}) with complexity $N^2\,D$,
    \item[$iii)$] the self-attention operation in (\ref{eq:self-attention-operation}) with complexity is $N^2\,D$,
    \item[$iv)$] the linear projection for the concatenated self-attention outputs in (\ref{eq:linear-projection-map-output}) with complexity $N\,D^2$. 
\end{itemize}
\noindent Summing the complexity of these steps yields the overall number of FLOPS for an MSA layer as $4\,N\,D^2 + 2\,N\,D^2$. Due to the quadratic dependence of the complexity on the sequence length $N$, AI researchers have been actively looking for novel and cheaper alternatives without sacrificing the MSA performance. The most promising of existing competitive methods is SSMs, especially the Mamba layer, which will be described in the next Section.
% While the multi-head attention layer has been serving as the backbone for almost all currently available LLMs
%The FLOPs mainly comes from four parts: (1) The projection of Q,K,V matrices φqkv(n, d) = 3nd2. (2) The calculation of the attention map φA(n, d) = n 2d. (3) The self-attention operation φO(n, d) = n 2d. (4) And finally, a linear projection for the concatenated self-attention outputs 

\subsection{State-space model layer}
In classical control and filtering theories, the evolution of continuous systems as a function of time $t$ with state $\bm{h}(t)\in\mathbb{R}^{D}$ and input $\bm{x}(t)\in\mathbb{R}^{N}$ is described according to the SSM: %describe the state of the system $\bm{h}(t)$ govern the evolution of the state $\bm{}$ by 
\begin{subequations}
\label{eq:state-space-model}
\begin{align}
    \bm{h}^\prime(t) &= \bm{A}\, \bm{h}(t) + \bm{B}\, \bm{x}(t),\hspace{1.55cm}\textrm{(state equation)} \\
    \bm{y}(t) &= \bm{C}\, \bm{h}(t) + \bm{D}\,\bm{x}(t),\hspace{1.5cm}\textrm{(output equation)}
\end{align}
\end{subequations}
%State space sequence models are a recent class of sequence-to-sequence models that are related to classical state space models as well as DL models such as RNNs, and CNNs.
\noindent In (\ref{eq:state-space-model}), the state equation describes how the state $\bm{h}(t)$ changes (through the matrix $\bm{A}$) based on how the input $\bm{x}(t)$ influences the state (through the matrix $\bm{B}$). The output equation describes how the state $\bm{h}(t)$ is observed in the output $\bm{y}(t) \in\mathbb{R}^{N}$ (through the matrix $\bm{C}$) and how the input $\bm{x}(t)$ influences the output (through the matrix $\bm{D}$). For sequence models, the input $\bm{x}(t)$ represents the token embedding at position $t$ while $\bm{y}(t)$ denotes the next token embedding. By learning the parameters $\bm{A}$, $\bm{B}$, $\bm{C}$ and $\bm{D}$, the SSM layer captures the evolution parameters of the dynamics from one token to the other. For systems with discrete state and input like textual sequences, the continuous-time SSM in (\ref{eq:state-space-model}) must be discretized using a step size $\boldsymbol{\Delta}$ which represents the resolution of the input. In other words, a discrete input $\bm{x}_t$ is a sample of the continuous input $\bm{x}(t)$ where $\bm{x}_t = \bm{x}(t\,\boldsymbol{\Delta})$. Using the bilinear method \cite{tustin1947method}, the discrete-time SSM is given by\footnote{Note that it is common to omit the parameter ${\bm{D}}$ during the discretization because the term ${\bm{D}}\,\bm{x}(t)$ is equivalent to a skip connection which be incorporated easily in the SSM layer architecture.}:
\begin{subequations}
\label{eq:state-space-model-discrete}
\begin{align}
    \bm{h}_t &= \xbar{\bm{A}}\, {\bm{h}}_{t-1} + \xbar{\bm{B}}\, \bm{x}_t,\label{eq:state-space-model-1}\\
    \bm{y}_t &= \xbar{\bm{C}}\, \bm{h}_t,\vspace{-0.5cm}% + \xbar{\bm{D}}\,\bm{x}_t,
\end{align}
\end{subequations}
where
\begin{subequations}\label{eq:state-space-model-discrete-variables}
    \begin{align}
        \xbar{\bm{A}} &~\triangleq~ (\bm{I}-\boldsymbol{\Delta} / 2 \cdot \bm{A})^{-1}(\bm{I}+\boldsymbol{\Delta} / 2 \cdot \bm{A}), \nonumber\\
        \xbar{\bm{B}} &~\triangleq~ (\bm{I}-\boldsymbol{\Delta} / 2 \cdot \bm{A})^{-1} \boldsymbol{\Delta}\, \bm{B},\\
        \xbar{\bm{C}} &~\triangleq~\bm{C}.
    \end{align}
\end{subequations}

\noindent The fact that the learnable parameters $\xbar{\bm{A}}$, $\xbar{\bm{B}}$, and $\xbar{\bm{C}}$ are constant means that the discrete-time SSM describes a linear time invariant (LTI) system with strong ties to convolution. Indeed, one can set the initial state $\bm{x}_{-1}$ to $\bm{0}$ for simplicity and rewrite (\ref{eq:state-space-model-discrete})--(\ref{eq:state-space-model-discrete-variables}) in the convolution representation for $t\in[1,T]$ as follows \cite{gu2021efficiently}:
\begin{equation}\label{eq:SSM-conv}
    \bm{y} = \xbar{\bm{K}}\ast\bm{x},
\end{equation}
where $\xbar{\bm{K}}\triangleq \left(\xbar{\boldsymbol{C}}\,\xbar{\boldsymbol{B}}, \xbar{\boldsymbol{C}}\,\xbar{\boldsymbol{A}}\,\xbar{\boldsymbol{B}}, \ldots, \xbar{\boldsymbol{C}}\,\overline{\boldsymbol{A}}^{T-1} \overline{\boldsymbol{B}}\right) $ represents the SSM convolution kernel. By avoiding the standard recurrent representation, the convolution representation in (\ref{eq:SSM-conv}) offers a compact and efficient parallel computation for SSM layers. However, because $\xbar{\bm{K}}$ is a giant filter, the naive implementation of the convolution as in (\ref{eq:SSM-conv}) is slow and memory inefficient.
To sidestep this limitation, many AI studies proposed restricting the structure of the SSM parameters to specific forms. Triangular $\xbar{\boldsymbol{A}}$ matrices keeping track of the Legendre polynomial's coefficients are computationally efficient and produce a hidden state $\bm{h}_t$ that memorizes the input history \cite{gu2020hippo}. Structured state space sequence models (S4) have also been introduced for SSMs where the parameters have a diagonal plus low-rank (DPLR) structure in the complex space \cite{gu2021efficiently}. Such a structure offers efficient SSMs with linear-time complexity instead of attention. More recently, Mamba \cite{gu2023mamba} enhanced the S4 model by introducing a selective input mechanism that enables the model to choose relevant information based on the input $\bm{x}_t$. This approach, combined with an implementation that is optimized for hardware, allowed Mamba to outperform transformers on different dense modalities like language and genomics. For input and state sequences $\bm{x}_t$ and $\bm{h}_t$ of size $N$ and $D$, the number of FLOPS of the Mamba layer scales linearly in $N$, more precisely it is $\mathcal{O}(N\,D)$. Another important aspect of the Mamba model is that it is the first SSM to be time-invariant by indirectly updating $\xbar{\bm{A}}$ through $\boldsymbol{\Delta}$ and directly updating $\xbar{\bm{B}}$ and $\xbar{\bm{C}}$ over time through its selective scan mechanism.

%modifying B and C to be selective allows finer-grained control over whether to let an input 푥

%all prior SSMs models must be time- and input-invariant in order to be computationally efficient.

%\textcolor{red}{talk about the non-LTI aspect of Mamba}

% for the complexity of Mamba see this answer
% https://github.com/state-spaces/mamba/issues/110#issuecomment-1919470069

% https://medium.com/aiguys/mamba-can-it-replace-transformers-fe2032537916
%\noindent The SSM enjoys fast inference (5× higher throughput than Transformers) and linear scaling in sequence length, and its performance improves on real data up to million-length sequences. Mamba achieves state-of-the-art performance across several modalities such as language, audio, and genomics as a general sequence model backbone. On language modeling, our Mamba-3B model outperforms Transformers of the same size and matches Transformers twice its size in pretraining and downstream evaluation.

\section{AI-based OFDM-CSI Prediction}\label{sec:prediction}
In this section, we describe the proposed CSI prediction mechanism for the next-slot OFDM-CSI prediction. We also describe how the input dimensions of SSM and MSA layers are mapped to the CSI dimensions. We then present the 3GPP channel models and their key parameters which will be later considered in our simulation results in Section \ref{sec:results}.
\subsection{Prediction tasks}
\begin{figure}[h!]
    \centering
    \includegraphics[scale=0.5]{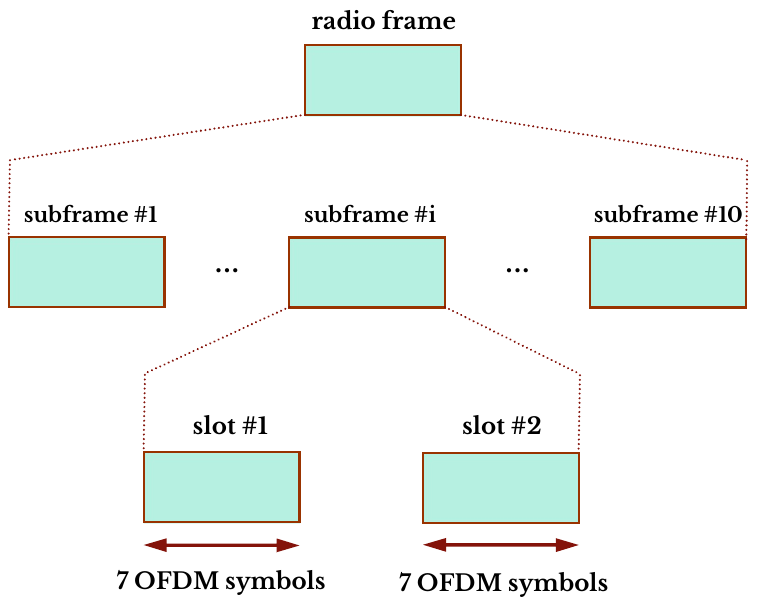}
    \caption{Illustration of the LTE radio frame structure}
    \label{fig:lte-frame-structure}
\end{figure}

In both 4G (a.k.a. LTE) and 5G networks, uplink and downlink transmissions are organized into radio frames of 10 ms each as depicted in Fig. \ref{fig:lte-frame-structure}. Each frame is divided into ten equally sized subframes. The duration of each subframe is 1 ms. In LTE, each subframe is further divided into two equal-size time slots, and each slot is of duration 0.5 ms. In 5G, however, the slot length changes depending on the used subcarrier spacing (a.k.a., numerology) associated with the operational frequency band and the service requirements.
%https://techlteworld.com/radio-frame-structure/#:~:text=However%2C%20unlike%20LTE%2C%20which%20has,of%20services%20and%20carrier%20frequencies.&text=LTE%20(4G)%3A,of%2015%20kHz%20is%20used.

\noindent In OFDM systems, the channel is a two-dimensional grid of $N_s$ symbols in time and $N_f$ sub-carriers in frequency. Specifically, consider a downlink system with $N_r$ antennas at the receiver (i.e., UE) and $N_t$ antennas at the transmitter (i.e., base station). The UE is continuously forecasting the CSI given the previously determined ones. To train and test SSM and MSA layers on this task, we consider the following CSI prediction problem: given the previous slot CSI, the UE predicts the CSI pertaining to next slot within the same subframe as depicted Fig. \ref{fig:lte-frame-structure}. This task covers slot-wise CSI prediction across subframes as well, i.e., between the last slot of subframe $i$ and the first slot of subframe $i+1$.%\footnote{Note that because the slot length in 5G is variable as a function of the numerology, directly predicting the next slot yields to a variable size prediction. In this scenario, one can generalize the next-slot prediction task to the next top-$k$ OFDM symbols' prediction task, which can yield directly predicting the next slot yields a variable number of forcasted OFDM symbols.}.

\begin{figure*}[b]
     \centering
     \begin{subfigure}[b]{0.49\textwidth}
         \centering
         \includegraphics[scale=0.41]{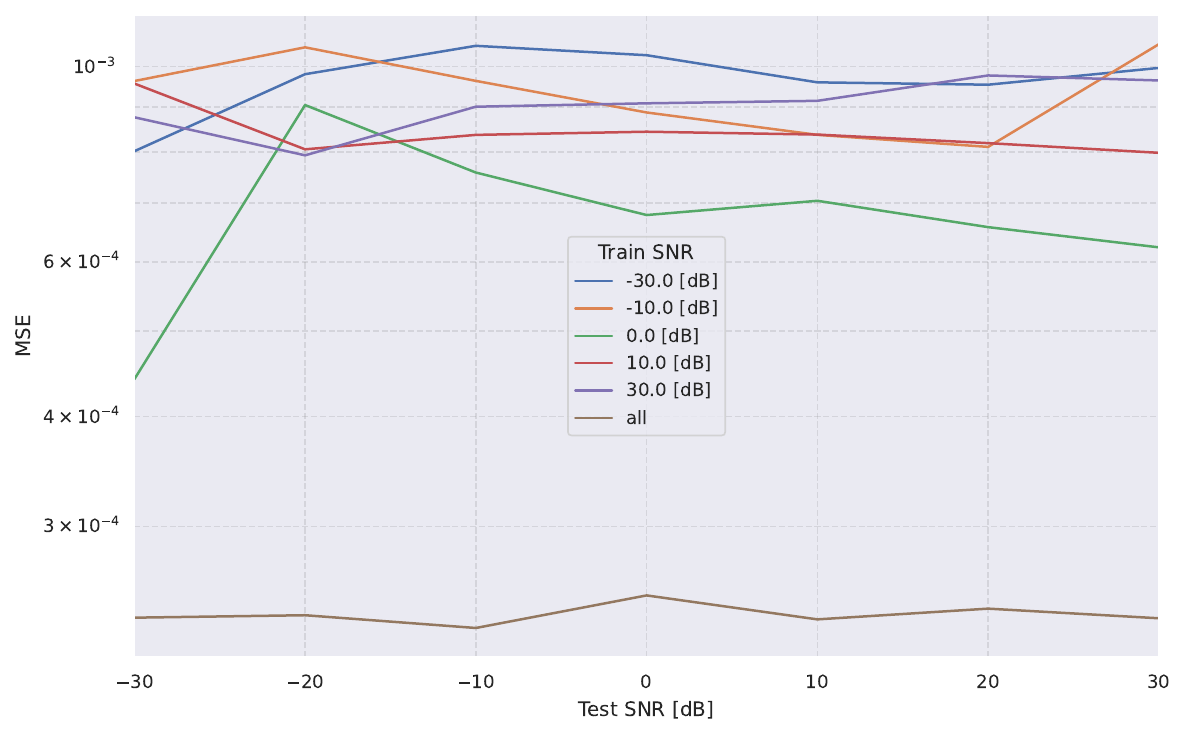}
         \caption{MSA, $v_{\textrm{train}} = 0$, $v_{\textrm{test}} = 0$}
         \label{fig:umi-slot-msa-trainV-0-testV-0}
     \end{subfigure}
     \begin{subfigure}[b]{0.49\textwidth}
         \centering
         \includegraphics[scale=0.41]{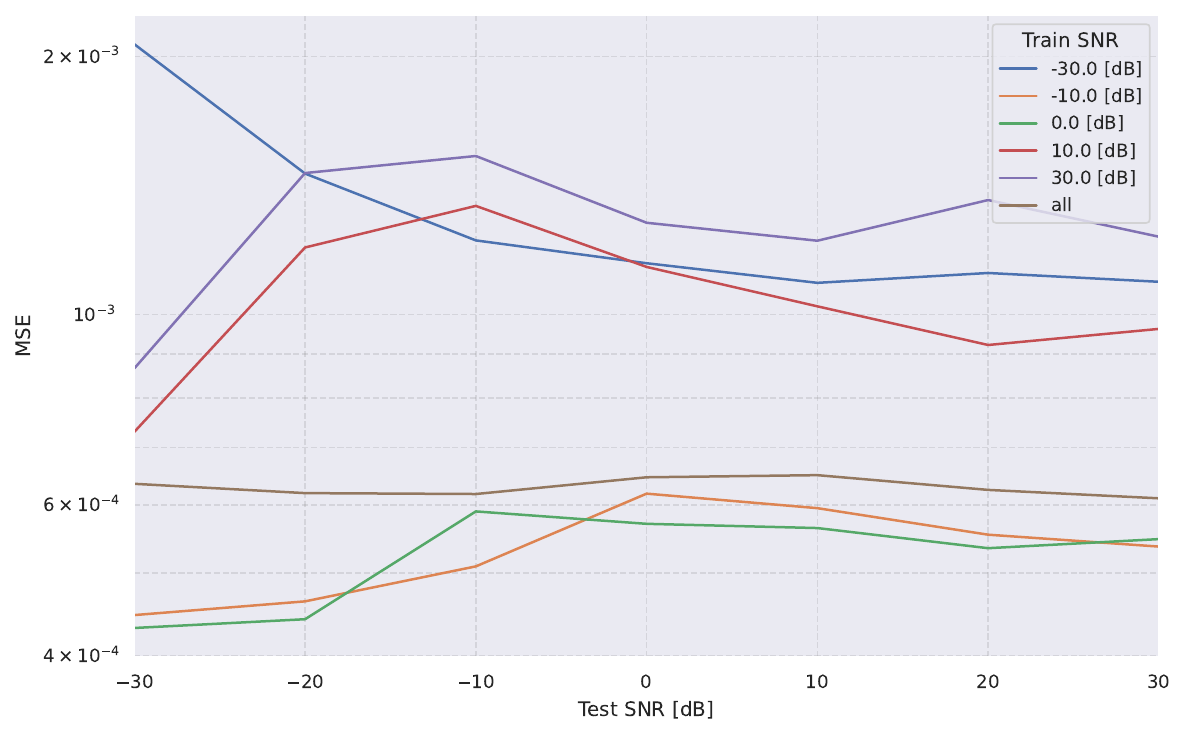}
         \caption{SSM, $v_{\textrm{train}} = 0$, $v_{\textrm{test}} = 0$}
         \label{fig:umi-slot-ssm-trainV-0-testV-0}
     \end{subfigure}
     \begin{subfigure}[b]{0.49\textwidth}
         \centering
         \includegraphics[scale=0.41]{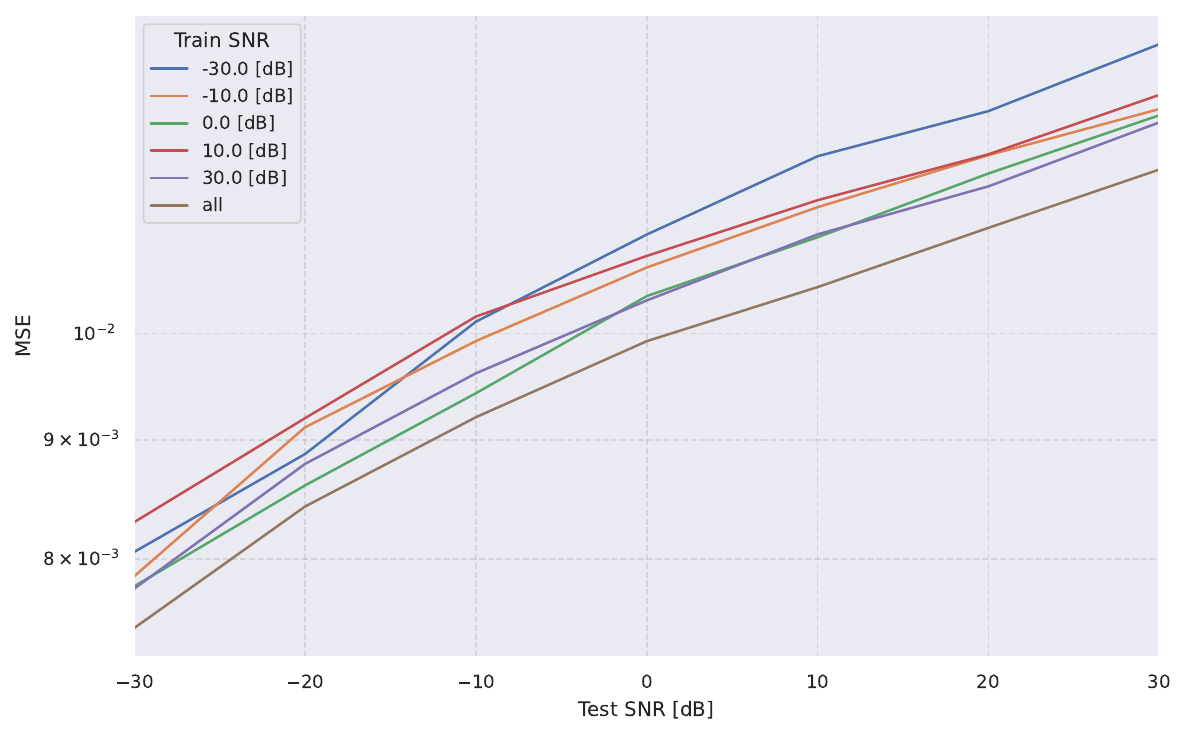}
         \caption{MSA, $v_{\textrm{train}} = 30$, $v_{\textrm{test}} = 30$}
         \label{fig:umi-slot-msa-trainV-30-testV-30}
     \end{subfigure}
     \begin{subfigure}[b]{0.49\textwidth}
         \centering
         \includegraphics[scale=0.41]{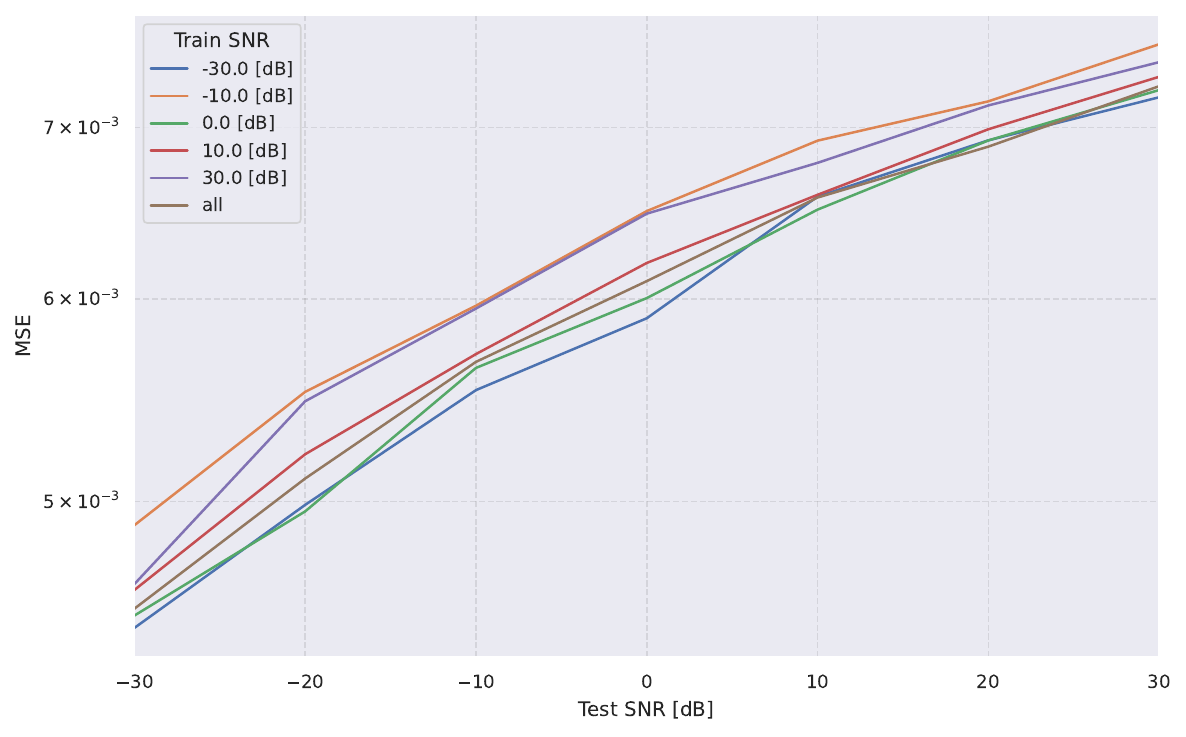}
         \caption{SSM, $v_{\textrm{train}} = 30$, $v_{\textrm{test}} = 30$}
         \label{fig:umi-slot-ssm-trainV-30-testV-30}
     \end{subfigure}
    \caption{SISO MSE of next-slot OFDM-CSI prediction task vs. test SNRs at $f_c=5$ GHz for multiple MSA layers in (a) and (c) and the SSM layers in (b) and (d) when each is trained with the UMi channel at different SNR values \textit{without a distribution shift} in the UE speed (i.e., $v_{\textrm{train}} = v_{\textrm{test}}$).}
    \label{fig:umi-slot-ID}
    %\vspace{-0.15cm}
\end{figure*}
Because the input of the tasks depends on the characteristics of the SSM and MSA layers, we associate their dimensions with those of the OFDM grid as follows:
\subsubsection{State-space model} Given $N_{s_0}$ OFDM symbols spanning $N_{f_0}$ sub-carriers, the input sequence $\bm{x}_t$ consists of $N_{s_0}$ symbols in time analogously to the token positions in sentences, while the sub-carrier dimension represents the number of channel coefficients in $\bm{x}_t$. As a result, the obtained input vector $\bm{x}_t$ belongs to $\mathbb{R}^{N_{s_0} \times 2 \,N_{f_0}}$, where the factor $2$ follows from the concatenation of the real and complex parts of the OFDM symbols.

\subsubsection{Multi-head attention} Similarly the SSM input, the $N_{s_0}$ OFDM symbols over the $N_{f_0}$ sub-carriers represent the input sequence of the attention layer. We use two attention heads for real and imaginary parts of the sequence.

\subsection{3GPP channel models}

We consider two 5G channel models from the 3GPP specification for frequency bands up to 100 GHz, namely the UMi and UMa channel models \cite{3gpp2017study}. They were derived based on extensive measurement and ray tracing results across a multitude of frequencies from 5 GHz to 100 GHz. We summarize in Table \ref{tab:channels-params} the key parameters we vary to assess the AI performance in next-slot OFDM-CSI prediction tasks.
\begin{table}[htbp]
  \centering
  \caption{Summary of the 3GPP channel parameters.}
    \begin{tabular}{cc}
        \toprule
          \textbf{Parameter} & \textbf{Values} \\
        \midrule
          OFDM channel type   & UMi, UMa         \\
          User speed [m/s]   & $\{0, 10, 20, 30\}$   \\
          SNR [dB]     & $\{-30,-10,0,10,30\}$ \\
          carrier frequency [GHz]     & \{5, 28\} \\
          carrier spacing [KHz]   & $30$ \\
          %modulation & 2-QAM\\
        \bottomrule
    \end{tabular}
  \label{tab:channels-params}
\end{table}

%The UMi, UMa, and RMa models require setting-up antenna models for the transmitters and receivers. This is achieved using the PanelArray class. The UMi, UMa, and RMa models require setting-up a network topology, specifying, e.g., the user terminals (UTs) and base stations (BSs) locations, the UTs velocities, etc. Utility functions are available to help laying out complex topologies or to quickly setup simple but widely used topologies.\\

%Urban microcell (UMi) channel model: Setting up a UMi model requires configuring the network topology, i.e., the UTs and BSs locations, UTs velocities, etc. This is achieved using the $set\_topology()$ method. Setting a different topology for each batch example is possible. The batch size used when setting up the network topology is used for the link simulations.\\

%Urban macrocell (UMa) channel model: Setting up a UMa model requires configuring the network topology, i.e., the UTs and BSs locations, UTs velocities, etc. This is achieved using the $set\_topology()$ method. Setting a different topology for each batch example is possible. The batch size used when setting up the network topology is used for the link simulations.\\

% page 24 of this https://www.etsi.org/deliver/etsi_tr/138900_138999/138901/16.01.00_60/tr_138901v160100p.pdf

\noindent Both UMi and UMa channels are considered without a line-of-sight between the base station and UEs.

\section{Numerical Results and Discussions}\label{sec:results}
In this section, we extensively assess the performance of SSM and MSA layers over multiple wireless scenarios. Throughout this section, we denote by $\mathcal{S}=\{-30,-10,0,10,30\}$ in [dB] and $\mathcal{V}=\{0,10,20,30\}$ in [m/s] the set of possible values for the SNR and user speeds. The code to train and test SSM and MSA layers is available on Github at {\scriptsize{\href{https://github.com/makrout/Next-Slot-OFDM-CSI-Prediction}{\textcolor{github-link}{\texttt{https://github.com/makrout/Next-Slot-OFDM-CSI-Prediction}}}}}.

\begin{figure*}[h]
     \centering
     \begin{subfigure}[b]{0.49\textwidth}
         \centering
         \includegraphics[scale=0.41]{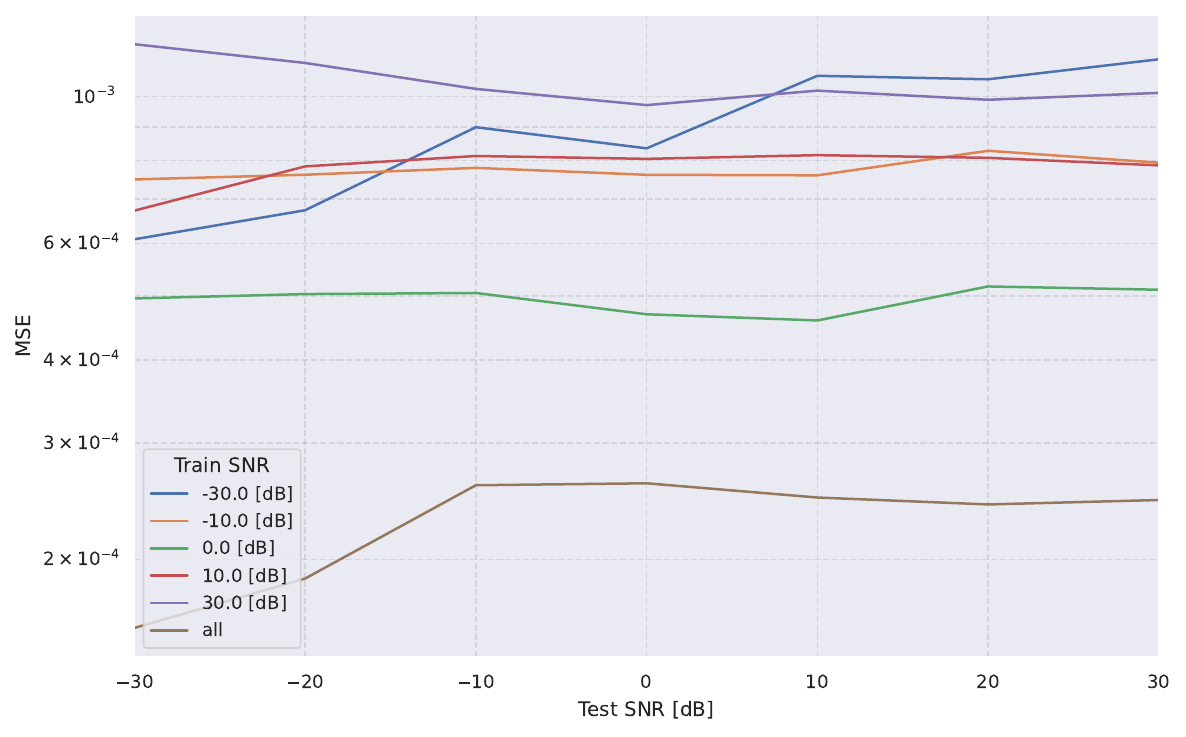}
         \caption{MSA, $v_{\textrm{train}} = 30$, $v_{\textrm{test}} = 0$}
         \label{fig:umi-slot-msa-trainV-30-testV-0}
     \end{subfigure}
     \begin{subfigure}[b]{0.49\textwidth}
         \centering
         \includegraphics[scale=0.41]{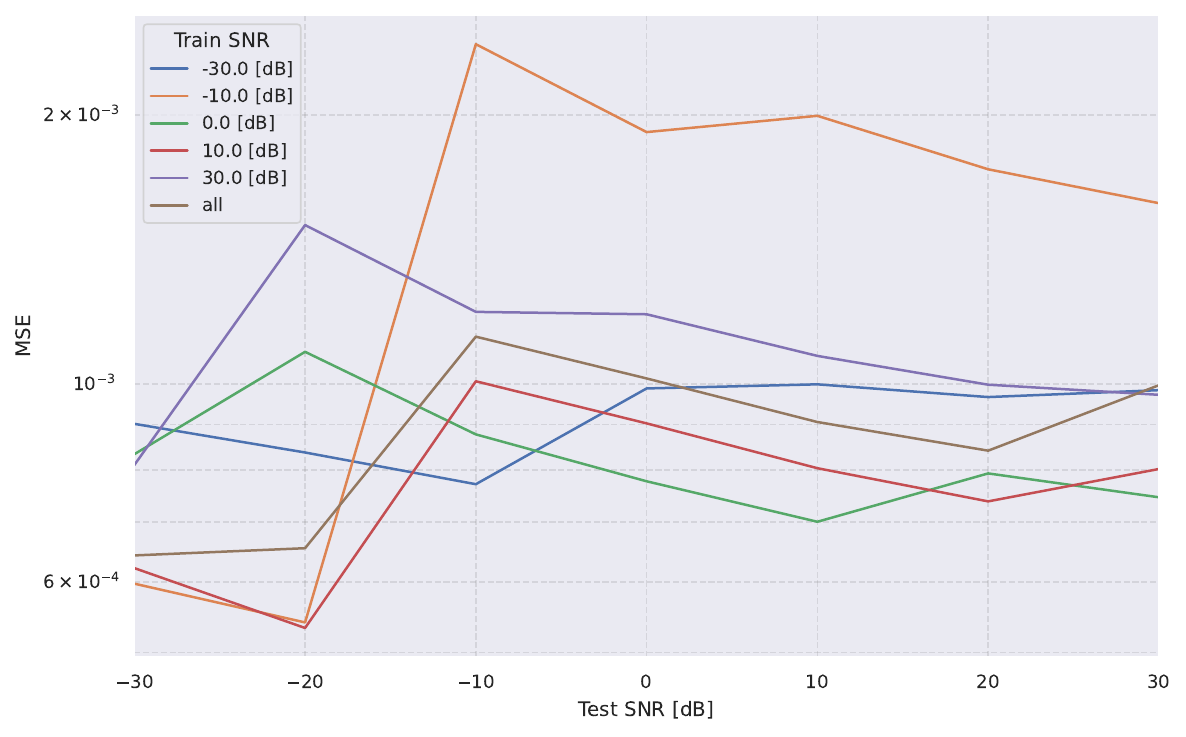}
         \caption{SSM, $v_{\textrm{train}} = 30$, $v_{\textrm{test}} = 0$}
         \label{fig:umi-slot-ssm-trainV-30-testV-0}
     \end{subfigure}
     \begin{subfigure}[b]{0.49\textwidth}
         \centering
         \vspace{0.2cm}
         \includegraphics[scale=0.41]{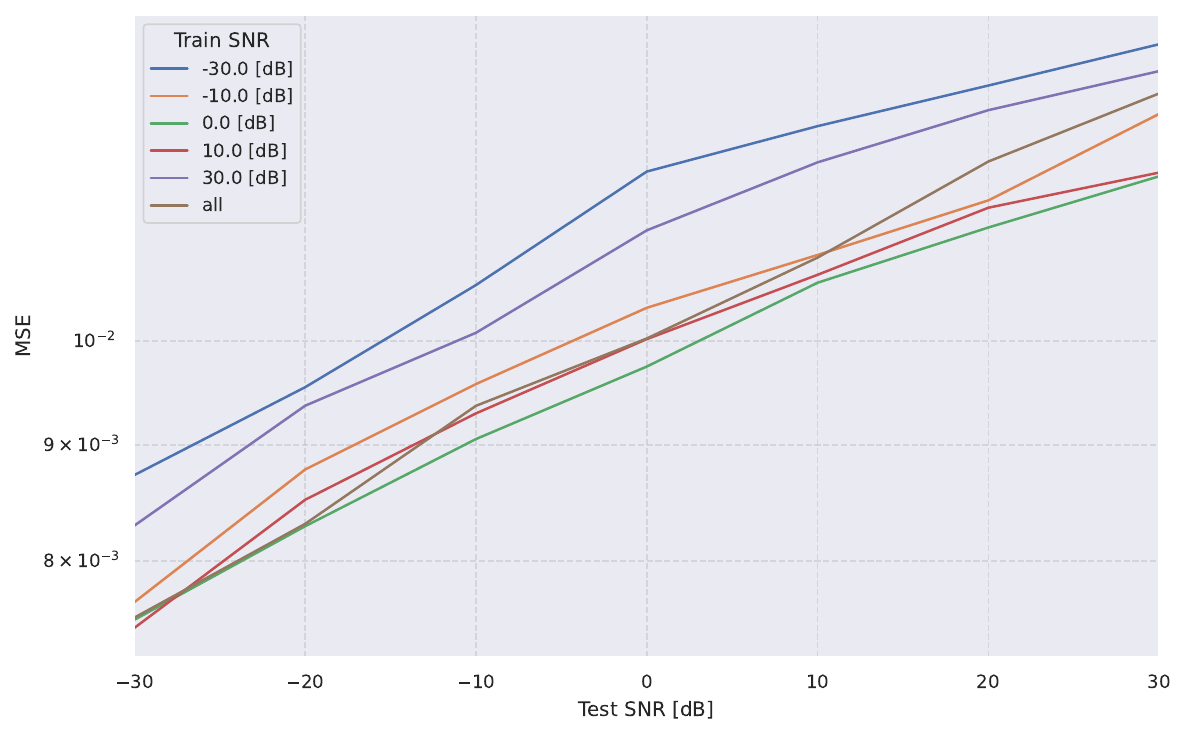}
         \caption{MSA, $v_{\textrm{train}} =0$, $v_{\textrm{test}} = 30$}
         \label{fig:umi-slot-msa-trainV-0-testV-30}
     \end{subfigure}
     \begin{subfigure}[b]{0.49\textwidth}
         \centering
         \includegraphics[scale=0.41]{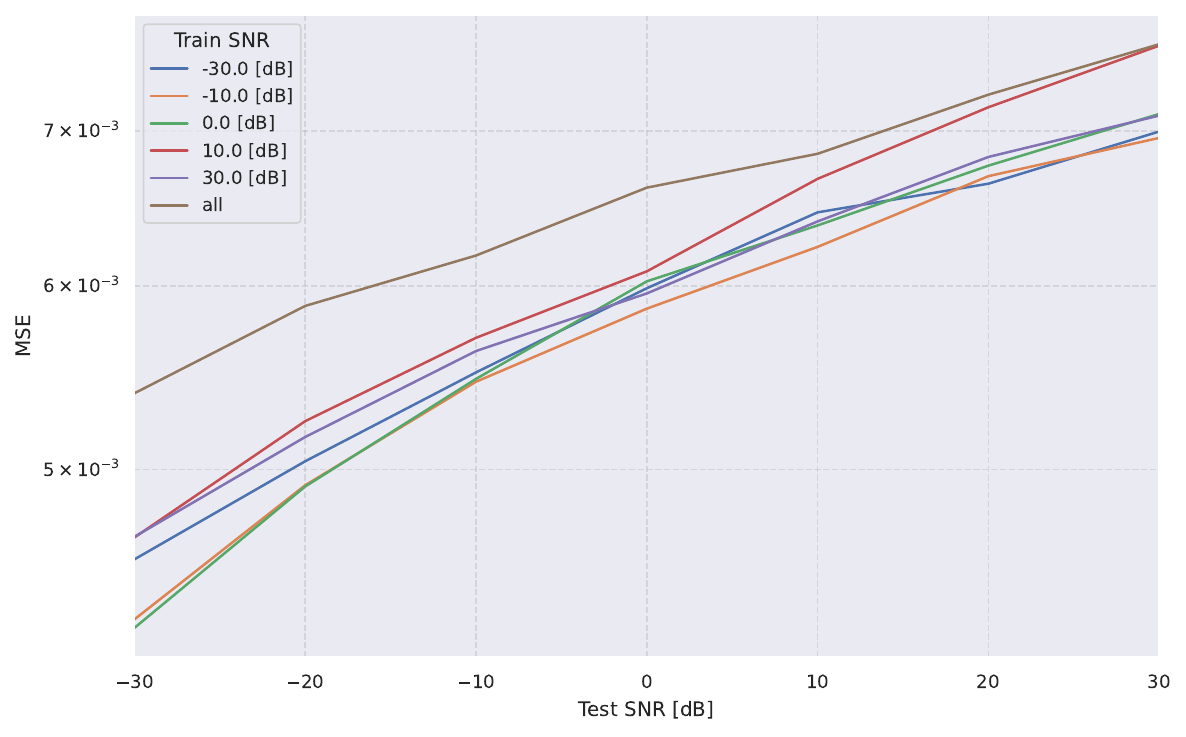}
         \caption{SSM, $v_{\textrm{train}} =0$, $v_{\textrm{test}} = 30$}
         \label{fig:umi-slot-ssm-trainV-0-testV-30}
     \end{subfigure}
    \caption{SISO MSE of next-slot OFDM-CSI prediction task vs. test SNRs  at $f_c=5$ GHz for multiple MSA layers in (a) and (c) and the SSM layers in (b) and (d) when each is trained with the UMi channel at different SNR values \textit{with a distribution shift} in the UE speed (i.e., $v_{\textrm{train}} \neq v_{\textrm{test}}$).}
    \label{fig:umi-slot-OOD-0-30}
    \vspace{-0.15cm}
\end{figure*}

\noindent Specifically, we consider the following two communication scenarios:
\begin{itemize}
    \item \textit{Uplink SISO transmission} between a base station and a single user, both having one antenna. The task on the user side forecasts the next-slot OFDM-CSI.
    \item \textit{Downlink MIMO transmission} between a base station with $n_{\textrm{T}_\textrm{x}}$ antennas and $n_\textrm{U}$ users each with one single antenna. The task at the base station side forecasts the next-slot OFDM-CSI prediction for all users simultaneously.
\end{itemize}
\noindent In all simulations, we consider transmissions using $2$-QAM constellations. For a fixed configuration of OFDM channel type, carrier frequency, and carrier spacing (cf. Table \ref{tab:channels-params}), we train MSA and SSM layers to minimize the MSE between the next-slot OFDM-CSI and the predicted one for a given communication scenario determined by the SNR and user speed values in $\mathcal{S}$ and $\mathcal{V}$, respectively. Users have single antennas with vertical polarization and an omnidirectional antenna pattern. The base station, however, has a uniform linear array with dual polarization, with each antenna element having a 3GPP 38.901 antenna pattern.

\noindent Then, we test the trained layers on the same speed values for ID evaluation and on different ones for OOD evaluation. We set the number of epochs to 1000 and we report the average MSE performance after 100 iterations. We use the Sionna library \cite{hoydis2022sionna} to generate the OFDM grids for each training and test communication scenario considered.

\noindent As mentioned in Section \ref{subsec:contrib}, we highlight the fact that the goal of our simulation results is not to design state-of-the-art DNN architectures for CSI predictions, but rather compare the predictive capability of the SSM and MSA layers only.

\subsection{SISO experiments}\label{subsec:SISO-simulations}
\subsubsection{In-distribution evaluation}\label{sec:ID-SISO}
For a fixed user speed $v_{\textrm{train}}$, we train separate MSA and SSM layers for each SNR level in $\mathcal{S}$ at $f_c=5$ GHz. We also train an additional MSA and SSM layers on a communication scenario over the UMi channel with all SNR levels combined by uniformly sampling the SNR over $\mathcal{S}$, which we refer to as the SNR value ``all''. Fig. \ref{fig:umi-slot-ID} depicts the OFDM-CSI prediction MSE pertaining to each considered network when it is  evaluated over all possible SNR levels for static and highly mobile users, i.e., $v_{\textrm{train}} = v_{\textrm{test}} \in \{0,30\}$. By comparing Figs. \ref{fig:umi-slot-msa-trainV-0-testV-0} and \ref{fig:umi-slot-ssm-trainV-0-testV-0}, it is seen that both SSM and MSA layers exhibit comparable MSE performance for static users only (i.e., $v=0$).
For mobile users with $v=30$, it is observed how SSMs in Fig. \ref{fig:umi-slot-ssm-trainV-30-testV-30} outperform MSAs in Fig. \ref{fig:umi-slot-msa-trainV-30-testV-30}, with the MSE being an order of magnitude smaller for SNR values larger than $0$ dB. However, the overall profile of the MSE over the entire SNR range increases for both models when users are mobile.% significantly increases the task complexity.

\begin{figure*}[b]
     \centering
     \begin{subfigure}[b]{0.49\textwidth}
         \centering
         \includegraphics[scale=0.41]{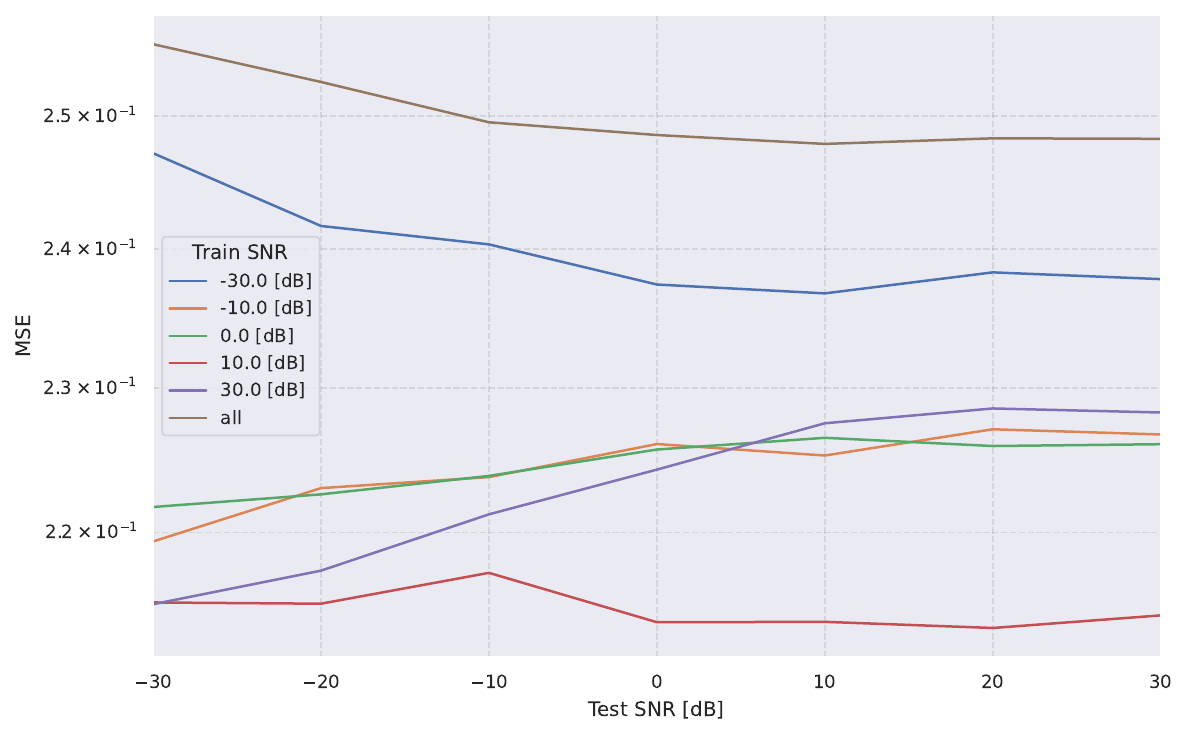}
         \caption{MSA, $v_{\textrm{train}} = 0$, $v_{\textrm{test}} = 0$}
         \label{fig:umi-slot-msa-trainV-0-testV-0-mimo}
     \end{subfigure}
     \begin{subfigure}[b]{0.49\textwidth}
         \centering
         \includegraphics[scale=0.41]{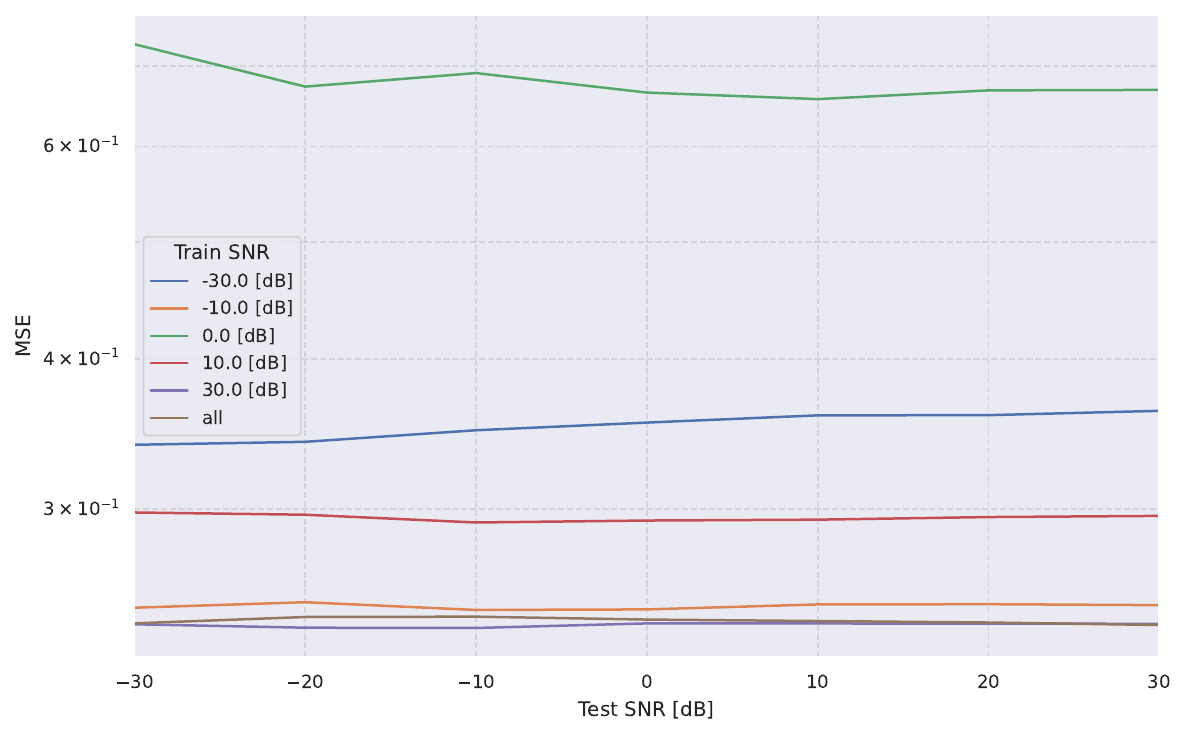}
         \caption{SSM, $v_{\textrm{train}} = 0$, $v_{\textrm{test}} = 0$}
         \label{fig:umi-slot-ssm-trainV-0-testV-0-mimo}
     \end{subfigure}
     \begin{subfigure}[b]{0.49\textwidth}
         \centering
         \includegraphics[scale=0.41]{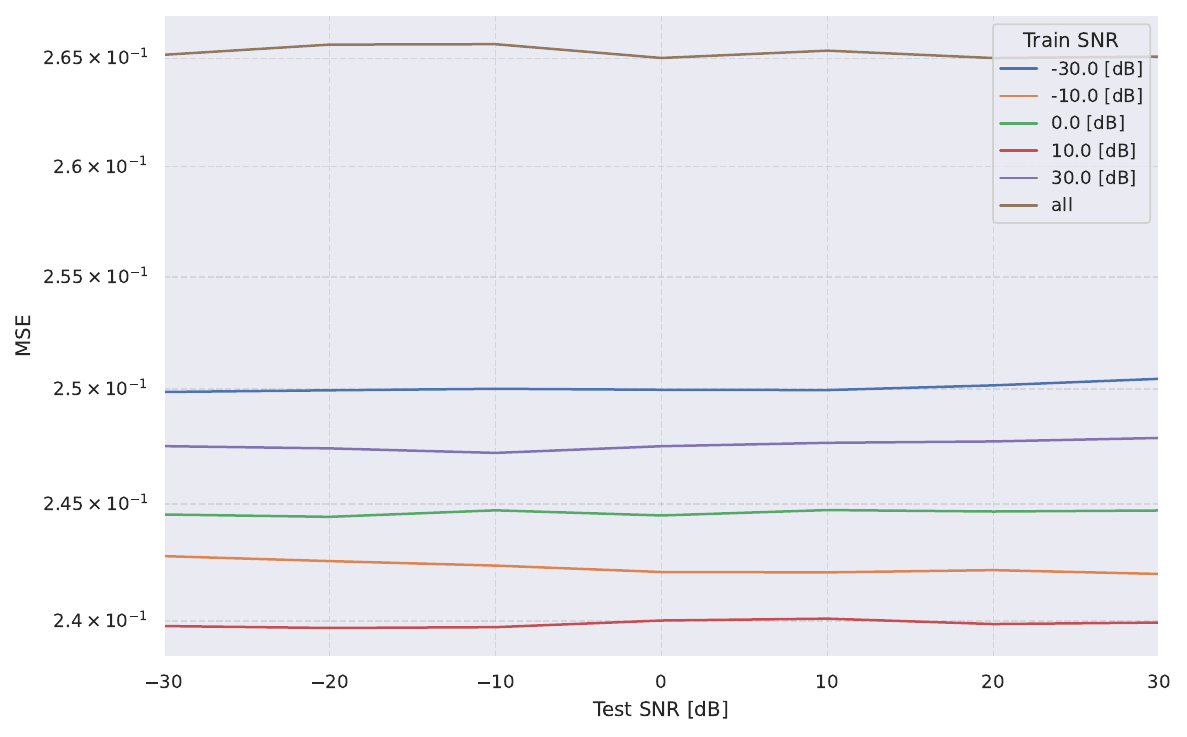}
         \caption{MSA, $v_{\textrm{train}} = 30$, $v_{\textrm{test}} = 30$}
         \label{fig:umi-slot-msa-trainV-30-testV-30-mimo}
     \end{subfigure}
     \begin{subfigure}[b]{0.49\textwidth}
         \centering
         \includegraphics[scale=0.41]{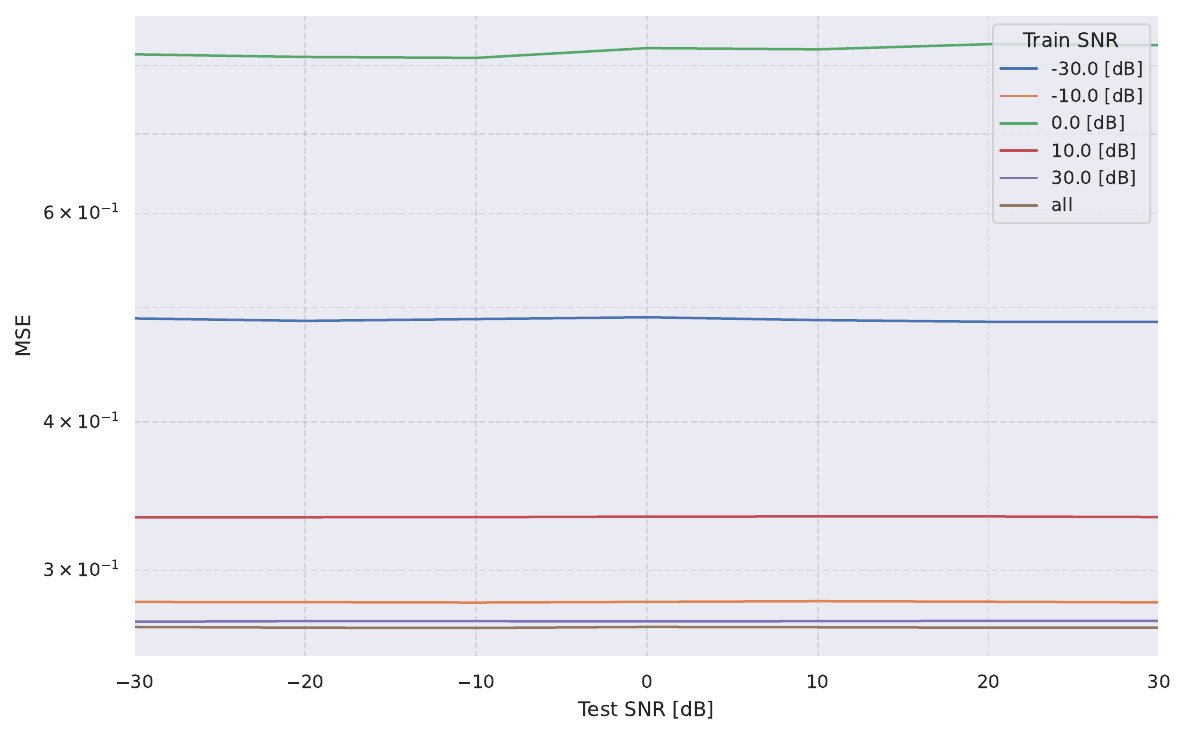}
         \caption{SSM, $v_{\textrm{train}} = 30$, $v_{\textrm{test}} = 30$}
         \label{fig:umi-slot-ssm-trainV-30-testV-30-mimo}
     \end{subfigure}
    \caption{MIMO MSE of next-slot OFDM-CSI prediction task vs. test SNRs at $f_c=5$ GHz for multiple MSA layers in (a) and (c) and the SSM layers in (b) and (d) when each is trained with the UMi channel at different SNR values \textit{without distribution shift} in the UE speed (i.e., $v_{\textrm{train}} = v_{\textrm{test}}$).}
    \label{fig:umi-slot-ID-mimo}
    %\vspace{-0.15cm}
\end{figure*}
\begin{figure*}[t!]
     \centering
     \begin{subfigure}[b]{0.49\textwidth}
         \centering
         \includegraphics[scale=0.41]{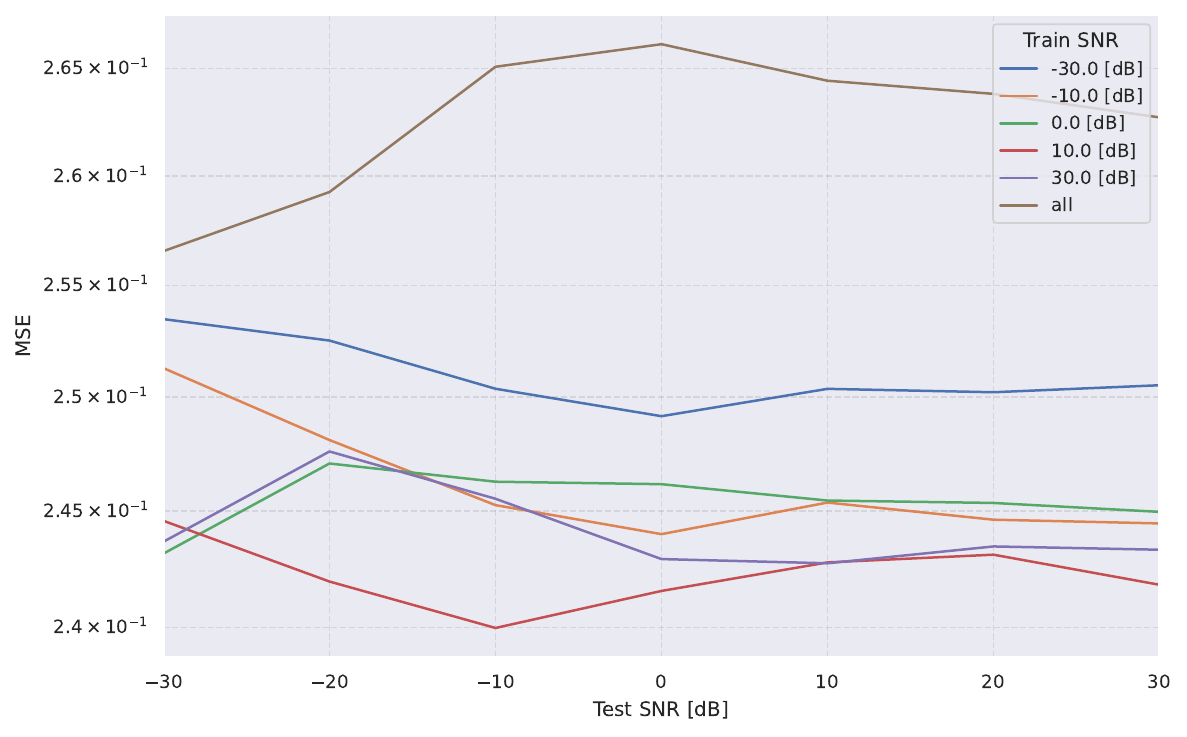}
         \caption{MSA, $v_{\textrm{train}} = 30$, $v_{\textrm{test}} = 0$}
         \label{fig:umi-slot-msa-trainV-30-testV-0-mimo}
     \end{subfigure}
     \begin{subfigure}[b]{0.49\textwidth}
         \centering
         \includegraphics[scale=0.41]{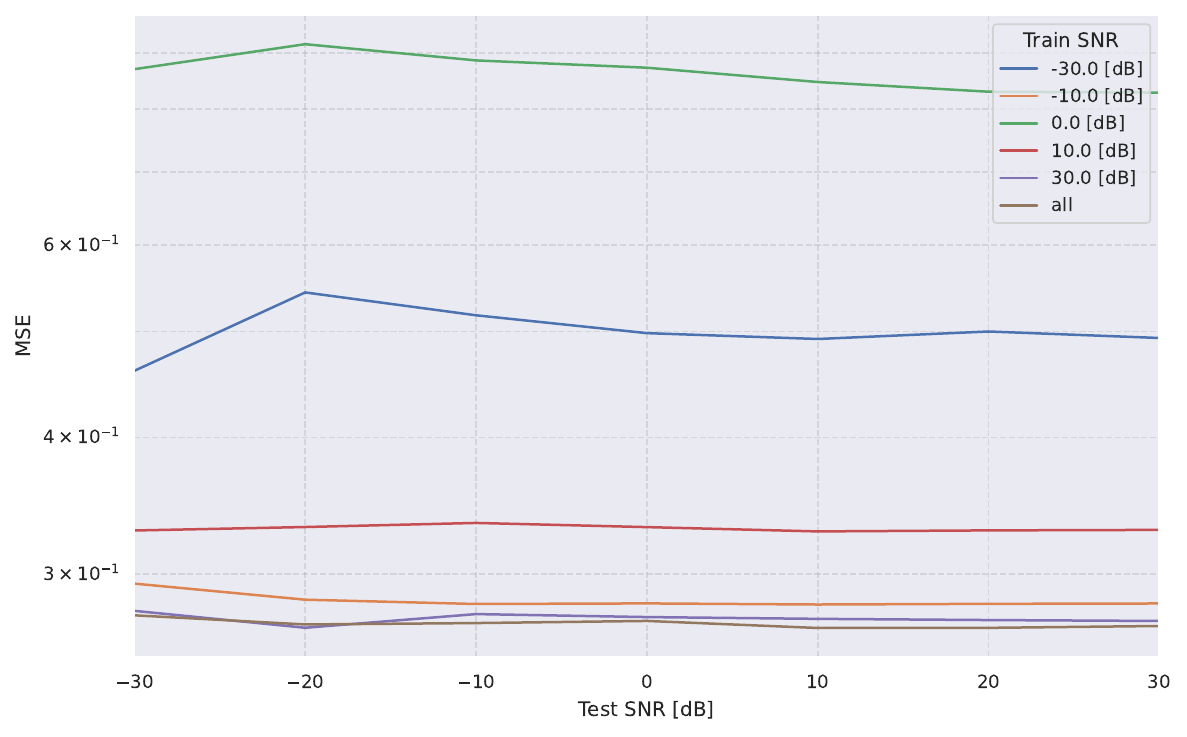}
         \caption{SSM, $v_{\textrm{train}} = 30$, $v_{\textrm{test}} = 0$}
         \label{fig:umi-slot-ssm-trainV-30-testV-0-mimo}
     \end{subfigure}
     \begin{subfigure}[b]{0.49\textwidth}
         \centering
         \vspace{0.2cm}
         \includegraphics[scale=0.41]{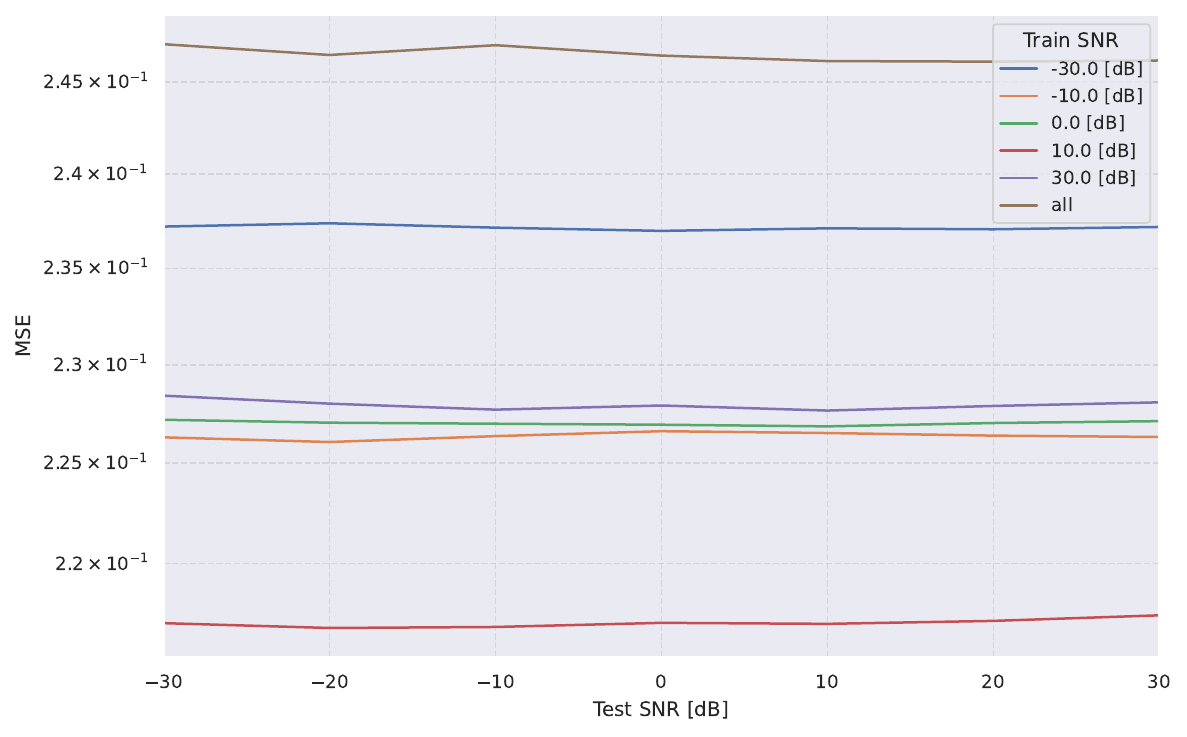}
         \caption{MSA, $v_{\textrm{train}} =0$, $v_{\textrm{test}} = 30$}
         \label{fig:umi-slot-msa-trainV-0-testV-30-mimo}
     \end{subfigure}
     \begin{subfigure}[b]{0.49\textwidth}
         \centering
         \includegraphics[scale=0.41]{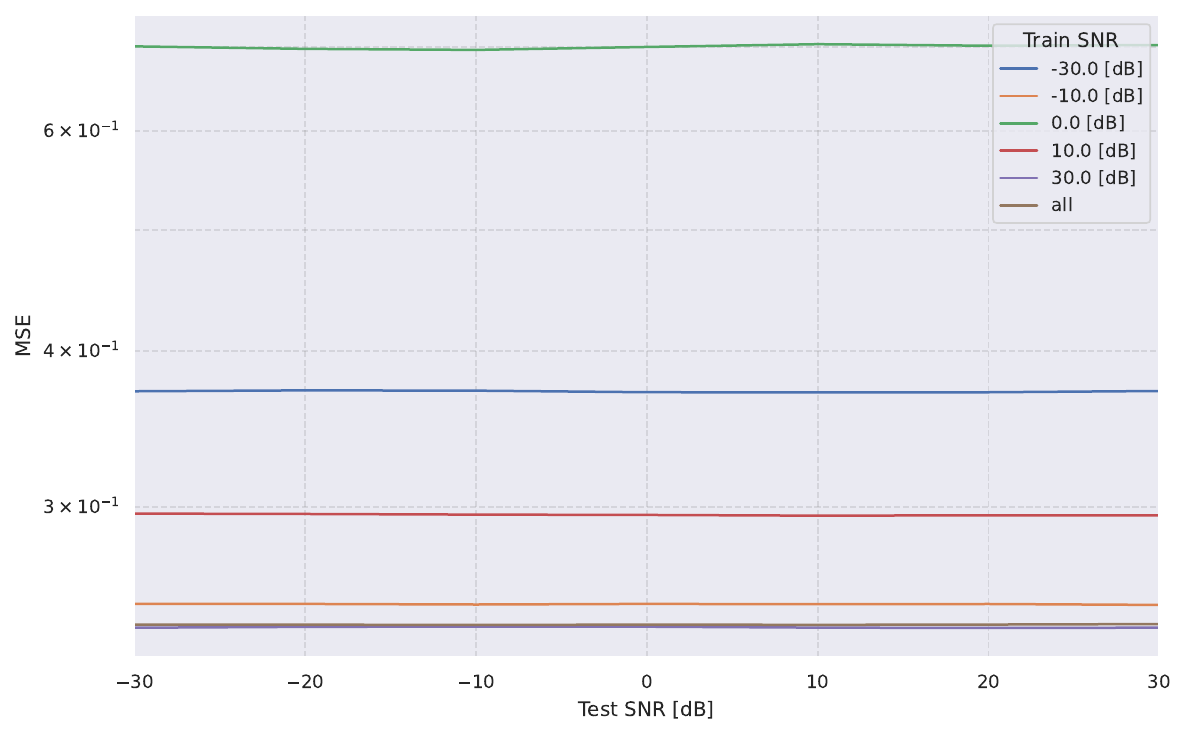}
         \caption{SSM, $v_{\textrm{train}} =0$, $v_{\textrm{test}} = 30$}
         \label{fig:umi-slot-ssm-trainV-0-testV-30-mimo}
     \end{subfigure}
    \caption{MIMO MSE of next-slot OFDM-CSI prediction task vs. test SNRs at $f_c=5$ GHz for multiple MSA layers in (a) and (c) and the SSM layers in (b) and (d) when each is trained with the UMi channel at different SNR values \textit{with a distribution shift} in the UE speed (i.e., $v_{\textrm{train}} \neq v_{\textrm{test}}$).}
    \label{fig:umi-slot-OOD-0-30-mimo}
    \vspace{-0.15cm}
\end{figure*}
\noindent On the other hand, Figs. \ref{fig:umi-slot-msa-trainV-0-testV-0} and \ref{fig:umi-slot-ssm-trainV-0-testV-0} show that the MSE decreases as a function of the SNR when users are static. This suggests that the task of learning the CSI prediction is more impacted by user mobility than the SNR. It is also noteworthy to observe how training the MSA layer with samples using all SNR levels (i.e., the gray curve) yields the lowest MSE across all test SNR levels. Interestingly, this SNR-wise diversification of samples, however, does not offer the lowest MSE for SSMs. This can be attributed to the fact that SSMs compress the signal sequence $\bm{x}$ in the state equation given in (\ref{eq:state-space-model-1}) by ensuring that $\bm{h}_t$ is a fixed-sized low-dimensional hidden state compared to $\bm{x}_t$. Such compression toward learning a state-space model or equivalently a transfer function\footnote{As a matter of fact, any state-space model can be seen as a transfer function in the Laplace domain \cite{leigh2004control}.} is not equally impacted by diversified communication scenarios in the dataset.

\noindent We then repeat the same training and evaluation at the mmwave carrier frequency $f_c = 28$ GHz. Simulation results are presented in Appendix \ref{appendix:f28-siso} due to space limitation where the ID and OOD MSE evaluations are reported in Figs. \ref{fig:umi-slot-ID-f25} and \ref{fig:umi-slot-OOD-0-30-f25}. There, similar MSE trends to those at $f_c=5$ GHz are observed. Overall, the MSE is higher due to the significant path loss in mmwave bands. It is also interesting to note that both SSMs and MSAs trained with all SNR values do not yield the lowest MSE performance. Similar MSE profiles reported in Appendix \ref{appendix:UMA-simulations} are also obtained after training and evaluating with UMa channels. The only notable difference is that MSE values are higher for UMa channels compared to UMi channels because macro-cell models cover a wider area with less dense networks.

\subsubsection{Out-of-distribution evaluation} Unlike the previous experiment, we now train and test MSAs and SSMs on different user speeds. In Figs. \ref{fig:umi-slot-msa-trainV-30-testV-0} and \ref{fig:umi-slot-ssm-trainV-30-testV-0}, we train on mobile user scenarios (i.e., $v_{\textrm{train}} = 30$) and test on static user scenarios (i.e., $v_{\textrm{test}} = 0$). We perform the inverse training and test strategy in Figs. \ref{fig:umi-slot-msa-trainV-0-testV-30} and \ref{fig:umi-slot-ssm-trainV-0-testV-30} by training on static users and testing on mobile ones. When comparing the range of the MSE between Figs. \ref{fig:umi-slot-msa-trainV-30-testV-0} and \ref{fig:umi-slot-ssm-trainV-30-testV-0} and Figs. \ref{fig:umi-slot-msa-trainV-0-testV-30} and \ref{fig:umi-slot-ssm-trainV-0-testV-30}, it is seen that training on challenging CSI prediction tasks (i.e., when users are mobile) and testing on easier ones (i.e., when users are static) provides a better MSE on OOD scenarios. We also note that networks trained on lower SNR values generalize better than those trained on higher SNR levels. Moreover, when testing with static users, SSMs and MSAs exhibit a similar range of MSEs as shown in Figs. \ref{fig:umi-slot-msa-trainV-30-testV-0} and \ref{fig:umi-slot-ssm-trainV-30-testV-0}. However, for test scenarios on mobile users, SSMs exhibit a much lower MSE profile compared to MSAs as shown in Figs. \ref{fig:umi-slot-msa-trainV-0-testV-30} and \ref{fig:umi-slot-ssm-trainV-0-testV-30}. Indeed, it is well known that adding noise to the training samples of a DNN is equivalent to the Tikhonov regularization and can lead to significant improvements in generalization performance \cite{bishop1995training}. Similarly to the ID evaluation in Section \ref{sec:ID-SISO}, the MSA model trained on all SNR levels is among the best performers for MSA layers, unlike the SSM ones.

\subsection{MIMO experiments}
For downlink MIMO CSI prediction, we consider a base station endowed with 20 transmit antennas communicating with 5 users, each of which with a single antenna.

%\textcolor{red}{add results then go back to previous sections and update with new insights}
\subsubsection{In-distribution evaluation}\label{subsec:ID-mimo}

Fig. \ref{fig:umi-slot-ID-mimo} shows the CSI prediction MSE of both SSM and MSA networks when evaluated over all possible SNR levels for static and mobile users, i.e., $v_{\textrm{train}} = v_{\textrm{test}} \in \{0,30\}$. Unlike the SISO case where the MSE performance of SSM and MSA layers were comparable, it is seen here that MSAs provide a lower MSE for both static and mobile user scenarios. It is interesting to observe again how networks trained with all SNR values do not provide the best performance. Training both models on MIMO scenarios with mobile or static users impacts the ID evaluation in the same way. This is to be opposed to the SISO case where training on mobile-user scenarios and testing on static-user scenarios yields better results than the opposite training and testing strategy. This does not reveal that the user speed is not a critical parameter but rather suggests that trained DNNs did not capture the correlation of fast-time varying channels due to the user mobility in MIMO scenarios. Indeed, the MSE is now two order of magnitude higher compared to the SISO case, suggesting that next-slot OFDM-CSI prediction is a challenging task for MIMO communication.

\subsubsection{Out-of-distribution evaluation}

%We train both MSA and SSM layers on MIMO scenarios with mobile or static users yields a very similar OOD MSE.

When we compare the MSE of mobile user training with static user evaluation against static user training with mobile user evaluation (i.e., Fig. \ref{fig:umi-slot-msa-trainV-30-testV-0-mimo} vs. Fig. \ref{fig:umi-slot-msa-trainV-0-testV-30-mimo} for MSAs, and Fig. \ref{fig:umi-slot-ssm-trainV-30-testV-0-mimo} and Fig. \ref{fig:umi-slot-ssm-trainV-0-testV-30-mimo} for SSMs), a negligible variation in the MSE is observed. This does not suggest that these models exhibit a strong generalization performance over user speeds given the high MSE values, but rather confirms the challenge in predicting the next-slot OFDM-CSI for MIMO scenarios as already reported for ID evaluation in Section \ref{subsec:ID-mimo}.

%\textcolor{red}{add Robert Heath comments + finish conclusion and appendices}

%\clearpage
\section{Conclusion}\label{sec:conclusion}

The existing applications of generative AI for wireless focus on language processing applications (e.g., prompt generation for compression, semantic communication). In this paper, we investigate the predictive capabilities of two key generative AI layers (i.e., multi-head attention and state space model) for OFDM slot prediction tasks. For these signal processing use cases, we compared the in-distribution and out-distribution performance of these two layers and empirically showed that multi-head attention layers outperform state space models for MIMO communication. However, we emphasize that the state space model layer has many advantages over the multi-head attention layer in terms of memory and computational complexity, which are also important factors for training and inference on long CSI inputs. Many avenues for further extension of this work are noteworthy. It is possible to design new hybrid architectures that endow state space models with an attention-like mechanism. One can also extend our benchmark to include more scenarios and models (e.g., antenna models, and frequency bands). One can also incorporate more wireless knowledge in the design of generative AI layers (e.g., prediction in the beam space).

\bibliographystyle{IEEEtran}
\bibliography{IEEEabrv,references}

%\vspace{1cm}

%\clearpage
\begin{appendices}
\renewcommand{\thesectiondis}[2]{\Roman{section}:}
\section{SISO Simulations at $f = 28$ GHz}\label{appendix:f28-siso}
\begin{figure}[h!]
     \centering
     \begin{subfigure}[b]{0.49\textwidth}
         \centering
         \includegraphics[scale=0.351]{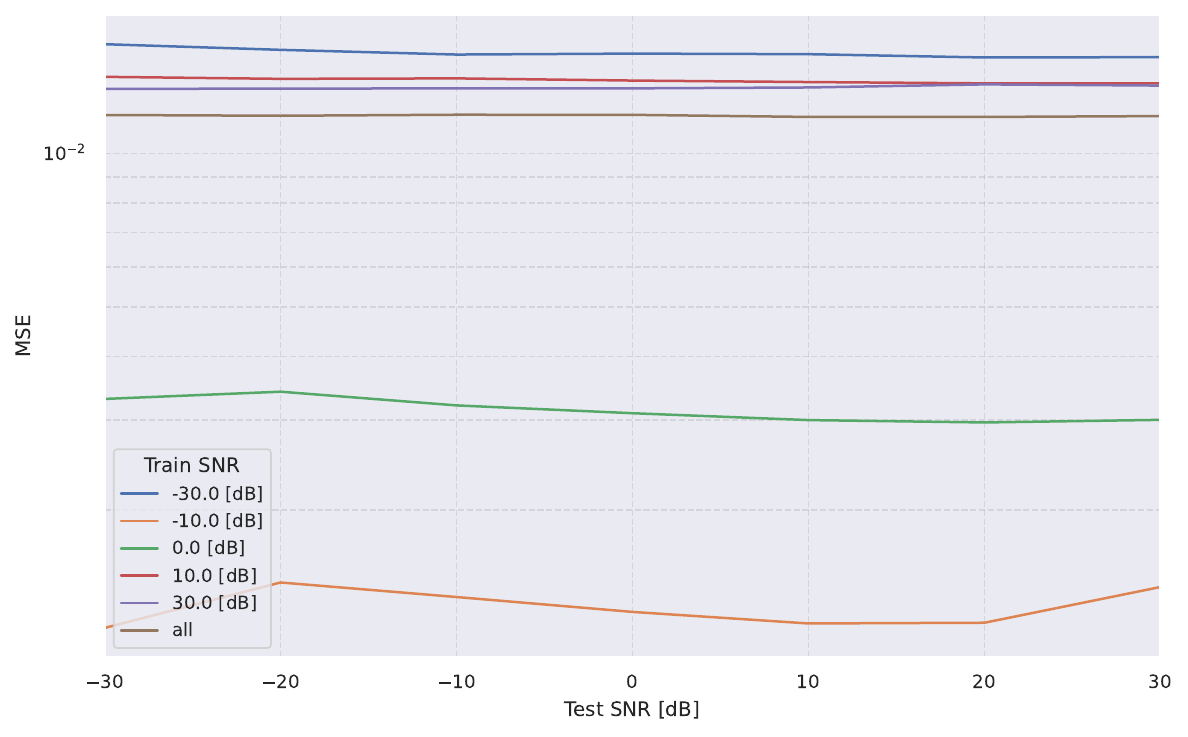}
         \caption{MSA, $v_{\textrm{train}} = 0$, $v_{\textrm{test}} = 0$}
         \label{fig:umi-slot-msa-trainV-0-testV-0-f25}
     \end{subfigure}
     %\vspace{0.1cm}
     \begin{subfigure}[b]{0.49\textwidth}
         \centering
         \includegraphics[scale=0.351]{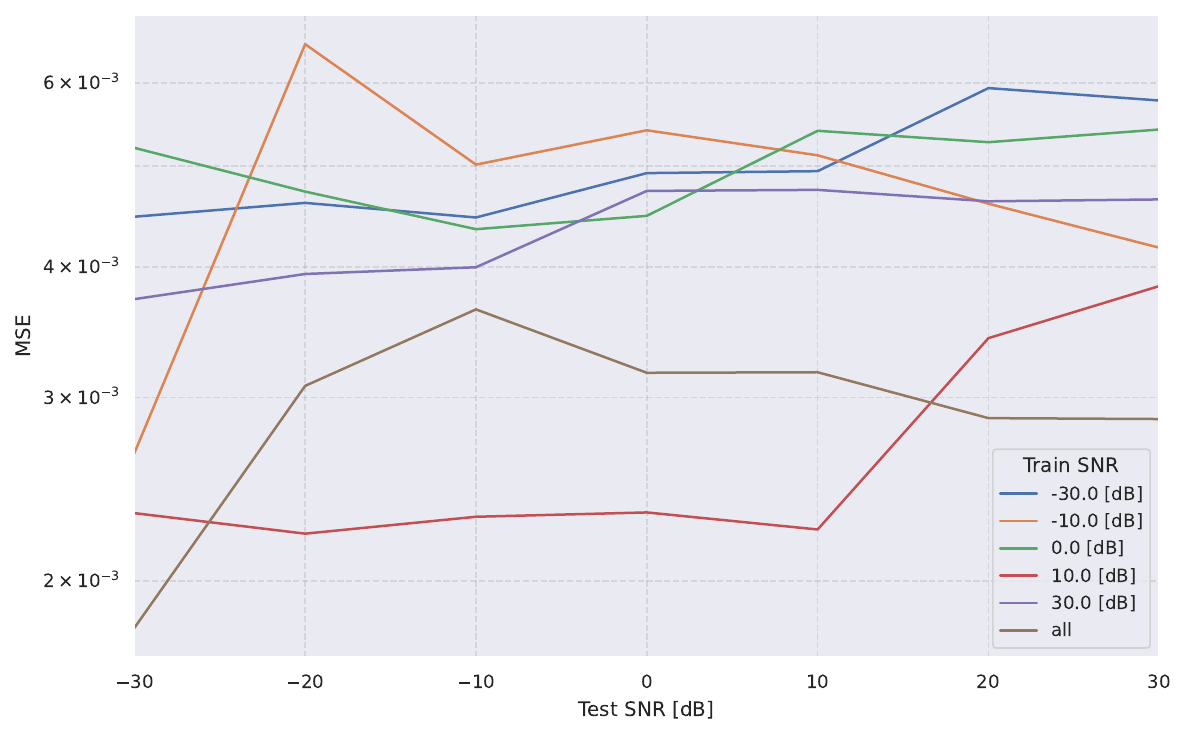}
         \caption{SSM, $v_{\textrm{train}} = 0$, $v_{\textrm{test}} = 0$}
         \label{fig:umi-slot-ssm-trainV-0-testV-0-f25}
     \end{subfigure}
     \begin{subfigure}[b]{0.49\textwidth}
         \centering
         \includegraphics[scale=0.351]{./figs/simulations/siso_1ue/f5GHz/mse_snr_msa_umi_slot_train_speed_30.0_test_speed_30.0.pdf}
         \caption{MSA, $v_{\textrm{train}} = 30$, $v_{\textrm{test}} = 30$}
         \label{fig:umi-slot-msa-trainV-30-testV-30-f25}
     \end{subfigure}
     \begin{subfigure}[b]{0.49\textwidth}
         \centering
         \includegraphics[scale=0.351]{./figs/simulations/siso_1ue/f5GHz/mse_snr_ssm_umi_slot_train_speed_30.0_test_speed_30.0.pdf}
         \caption{SSM, $v_{\textrm{train}} = 30$, $v_{\textrm{test}} = 30$}
         \label{fig:umi-slot-ssm-trainV-30-testV-30-f25}
     \end{subfigure}
    \caption{SISO MSE of next-slot OFDM-CSI prediction task vs. test SNRs at $f_c=28$ GHz for multiple MSA layers in (a) and (c) and the SSM layers in (b) and (d) when each is trained with the UMi channel at different SNR values \textit{without distribution shift} in the UE speed (i.e., $v_{\textrm{train}} = v_{\textrm{test}}$).}
    \label{fig:umi-slot-ID-f25}
    %\vspace{0.5cm}
\end{figure}

\begin{figure}[!htp]
     \centering
     \begin{subfigure}[b]{0.49\textwidth}
         \centering
         \includegraphics[scale=0.36]{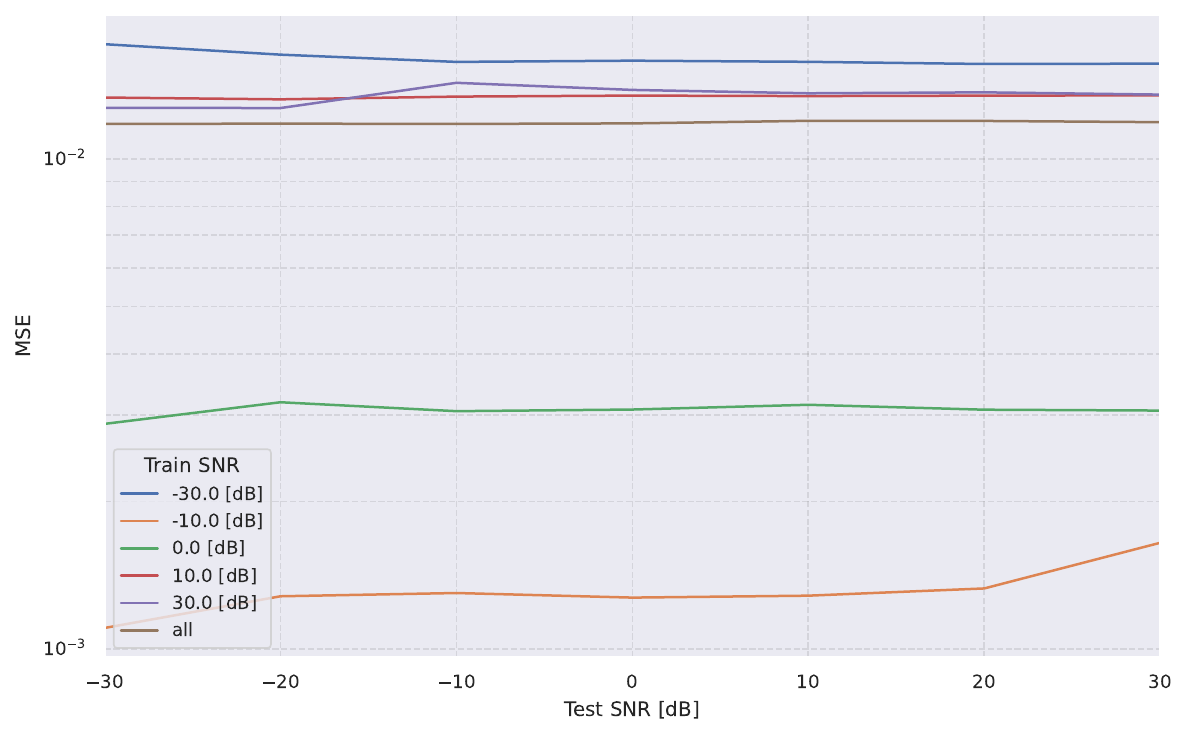}
         \caption{MSA, $v_{\textrm{train}} = 30$, $v_{\textrm{test}} = 0$}
         \label{fig:umi-slot-msa-trainV-30-testV-0-f25}
     \end{subfigure}
     \begin{subfigure}[b]{0.49\textwidth}
         \centering
         \includegraphics[scale=0.36]{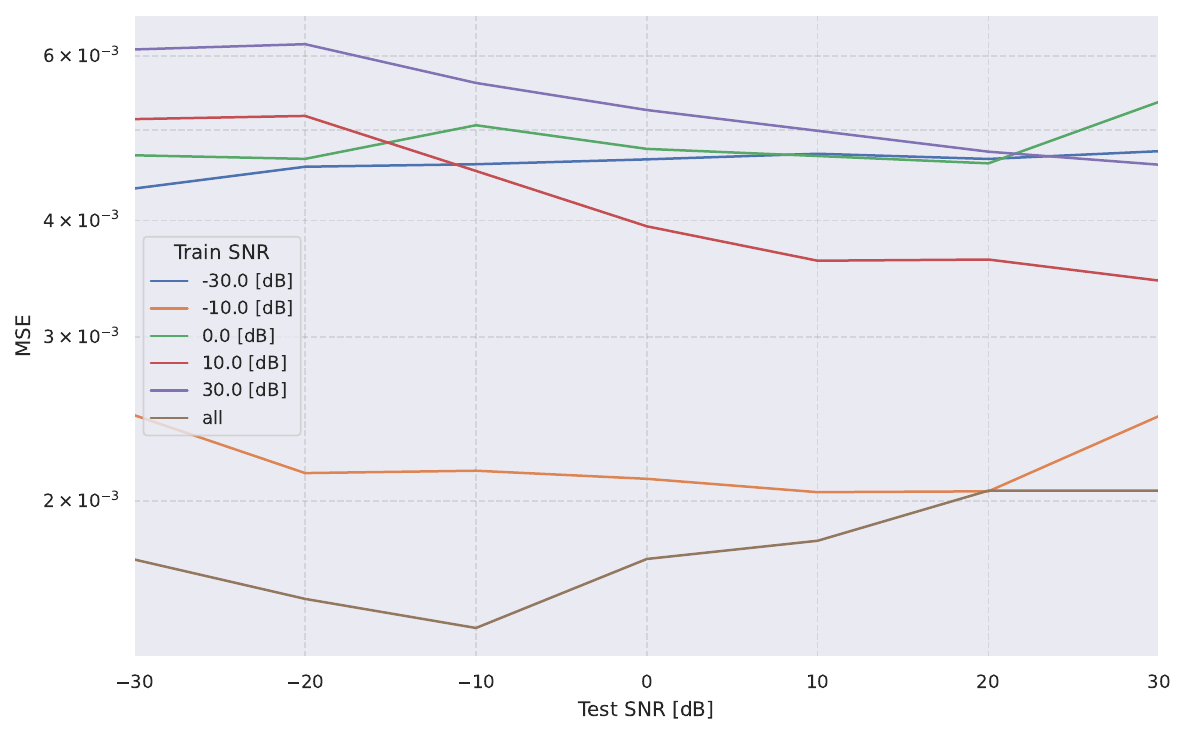}
         \caption{SSM, $v_{\textrm{train}} = 30$, $v_{\textrm{test}} = 0$}
         \label{fig:umi-slot-ssm-trainV-30-testV-0-f25}
     \end{subfigure}
     \begin{subfigure}[b]{0.49\textwidth}
         \centering
         \vspace{0.2cm}
         \includegraphics[scale=0.36]{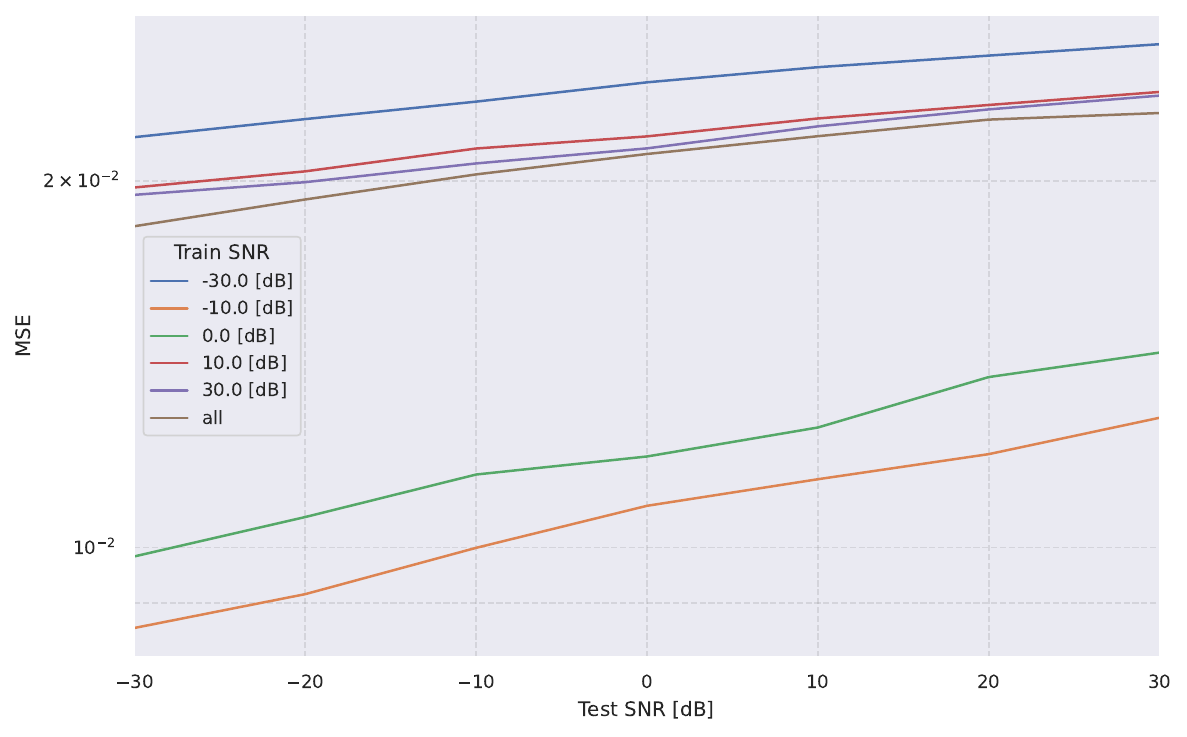}
         \caption{MSA, $v_{\textrm{train}} =0$, $v_{\textrm{test}} = 30$}
         \label{fig:umi-slot-msa-trainV-0-testV-30-f25}
     \end{subfigure}
     \begin{subfigure}[b]{0.49\textwidth}
         \centering
         \includegraphics[scale=0.36]{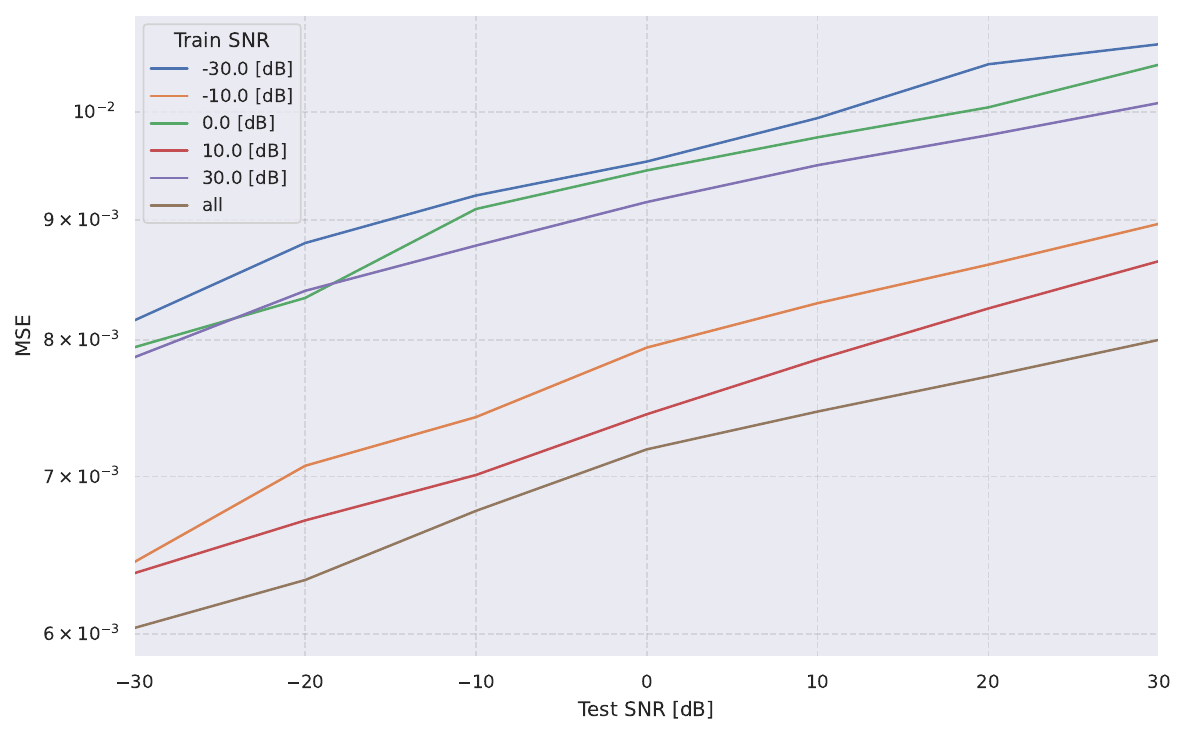}
         \caption{SSM, $v_{\textrm{train}} =0$, $v_{\textrm{test}} = 30$}
         \label{fig:umi-slot-ssm-trainV-0-testV-30-f25}
     \end{subfigure}
    \caption{SISO MSE of next-slot OFDM-CSI prediction task vs. test SNRs at $f_c=28$ GHz for multiple MSA layers in (a) and (c) and the SSM layers in (b) and (d) when each is trained with the UMi channel at different SNR values \textit{with a distribution shift} in the UE speed (i.e., $v_{\textrm{train}} \neq v_{\textrm{test}}$).}
    \label{fig:umi-slot-OOD-0-30-f25}
    %\vspace{1cm}
\end{figure}

In this appendix, we present the evaluation of both SSM and MSA layers for the SISO communication scenario described in Section \ref{subsec:SISO-simulations} when the carrier frequency is fixed at $f_c=28$ GHz. Figs. \ref{fig:umi-slot-ID-f25} and \ref{fig:umi-slot-OOD-0-30-f25} depict the ID and OOD evaluations.

%\clearpage

%\vspace{0.3cm}

\section{Simulations with the UMa channel}\label{appendix:UMA-simulations}
In this appendix, we present the results when SSM and MSA layers are trained on UMa channels.
\begin{figure}[!htp]
     \centering
     \begin{subfigure}[b]{0.49\textwidth}
         \centering
         \includegraphics[scale=0.36]{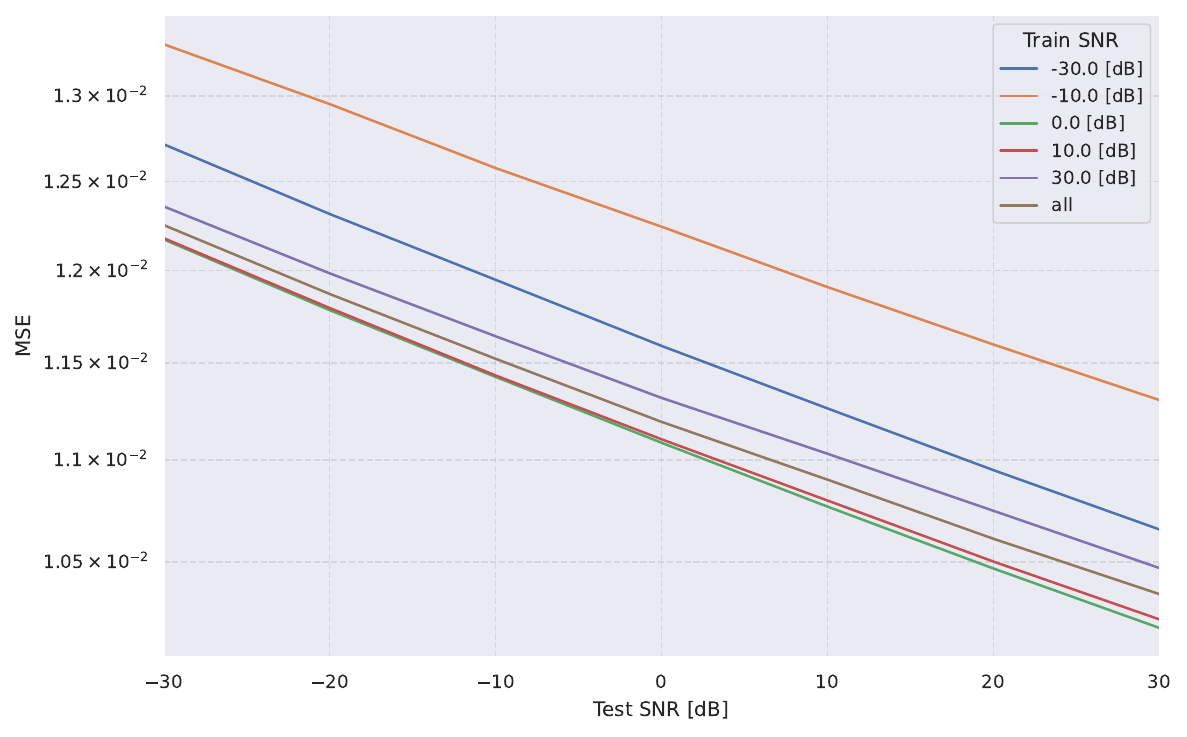}
         \caption{MSA, $v_{\textrm{train}} = 0$, $v_{\textrm{test}} = 0$}
         \label{fig:uma-slot-msa-trainV-0-testV-0}
     \end{subfigure}
     \begin{subfigure}[b]{0.49\textwidth}
         \centering
         \includegraphics[scale=0.36]{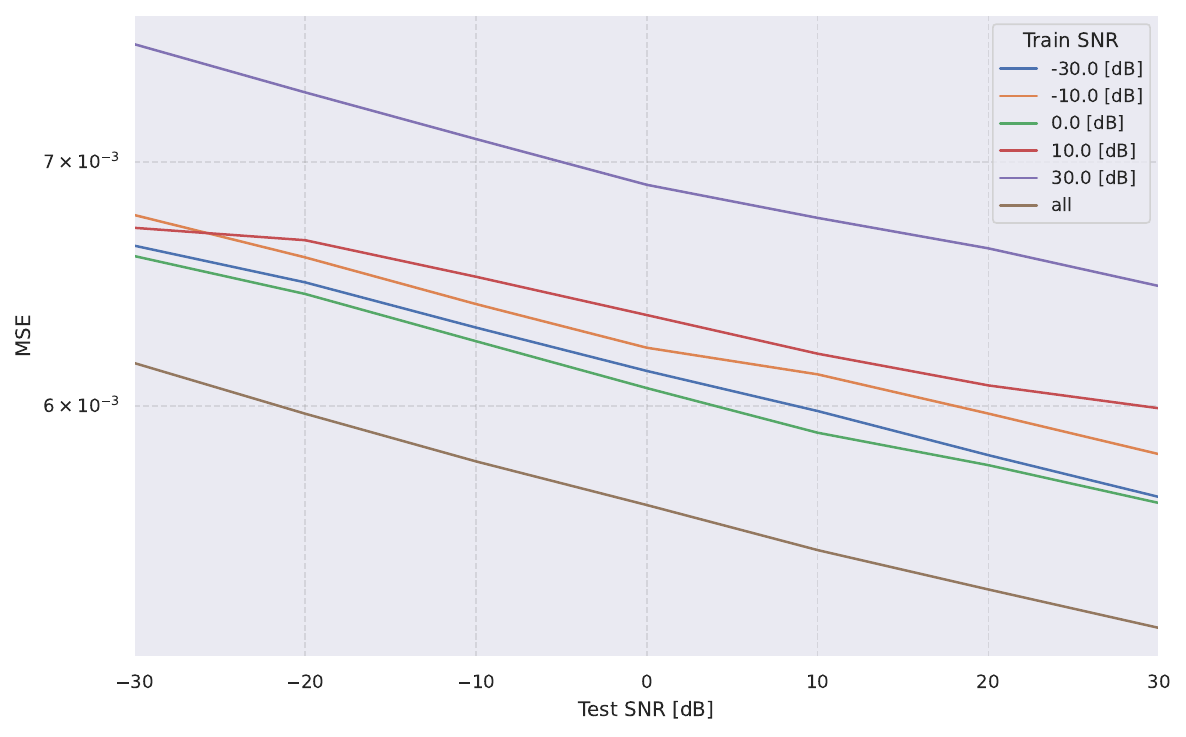}
         \caption{SSM, $v_{\textrm{train}} = 0$, $v_{\textrm{test}} = 0$}
         \label{fig:uma-slot-ssm-trainV-0-testV-0}
     \end{subfigure}
     \begin{subfigure}[b]{0.49\textwidth}
         \centering
         \includegraphics[scale=0.36]{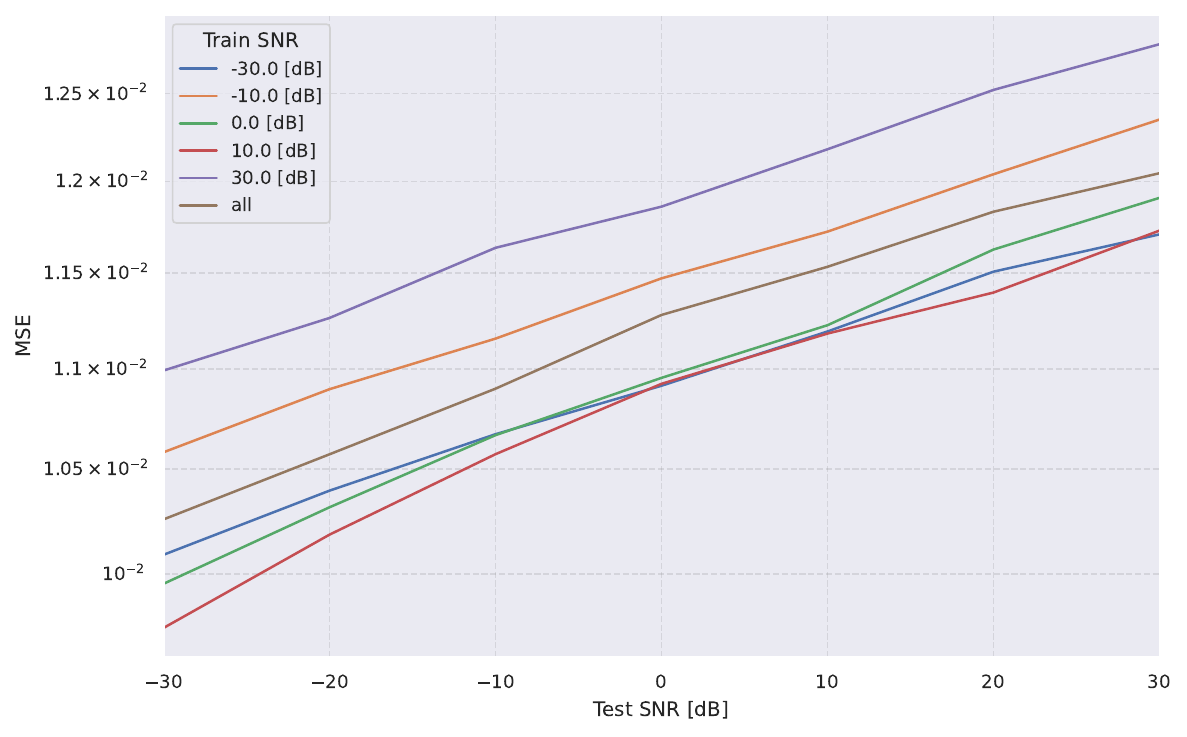}
         \caption{MSA, $v_{\textrm{train}} = 30$, $v_{\textrm{test}} = 30$}
         \label{fig:uma-slot-msa-trainV-30-testV-30}
     \end{subfigure}
     \begin{subfigure}[b]{0.49\textwidth}
         \centering
         \includegraphics[scale=0.36]{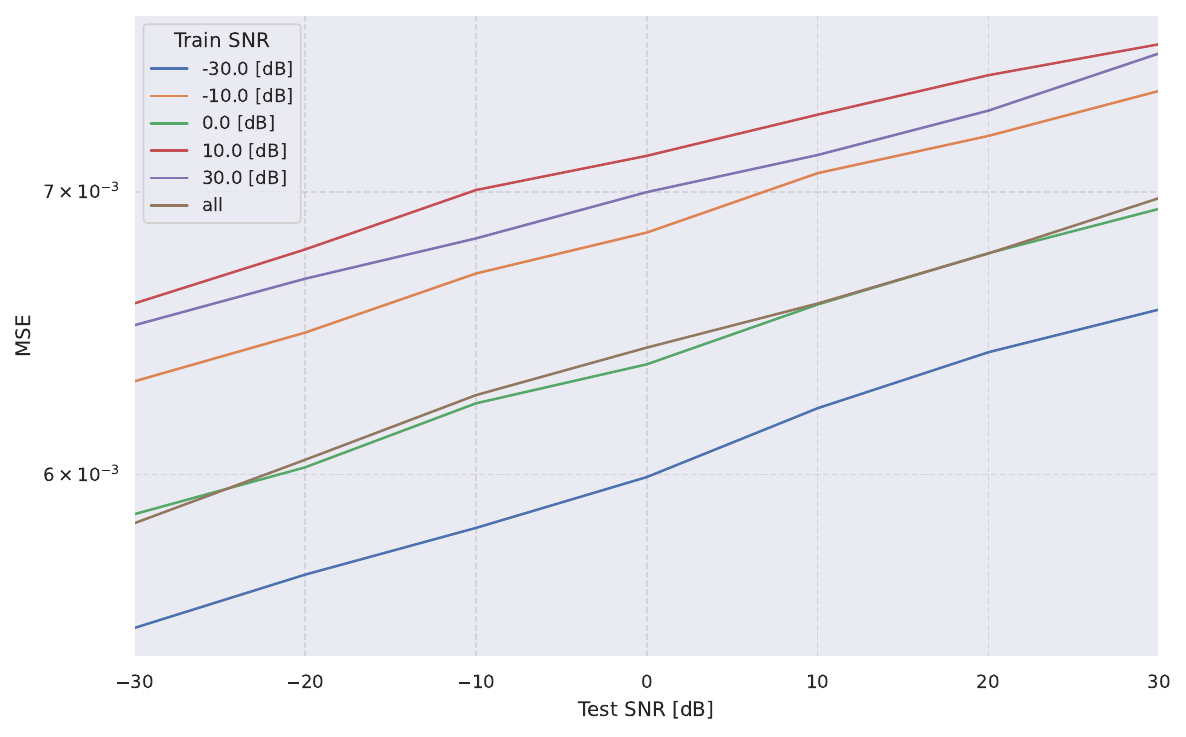}
         \caption{SSM, $v_{\textrm{train}} = 30$, $v_{\textrm{test}} = 30$}
         \label{fig:uma-slot-ssm-trainV-30-testV-30}
     \end{subfigure}
    \caption{SISO MSE of next-slot OFDM-CSI prediction task vs. test SNRs at $f_c=5$ GHz for multiple MSA layers in (a) and (c) and the SSM layers in (b) and (d) when each is trained with the UMa channel at different SNR values \textit{without distribution shift} in the UE speed (i.e., $v_{\textrm{train}} = v_{\textrm{test}}$).}
    \label{fig:uma-slot-ID}
    %\vspace{-0.15cm}
\end{figure}
\begin{figure}%[!htp]
     \centering
     \begin{subfigure}[b]{0.49\textwidth}
         \centering
         \includegraphics[scale=0.37]{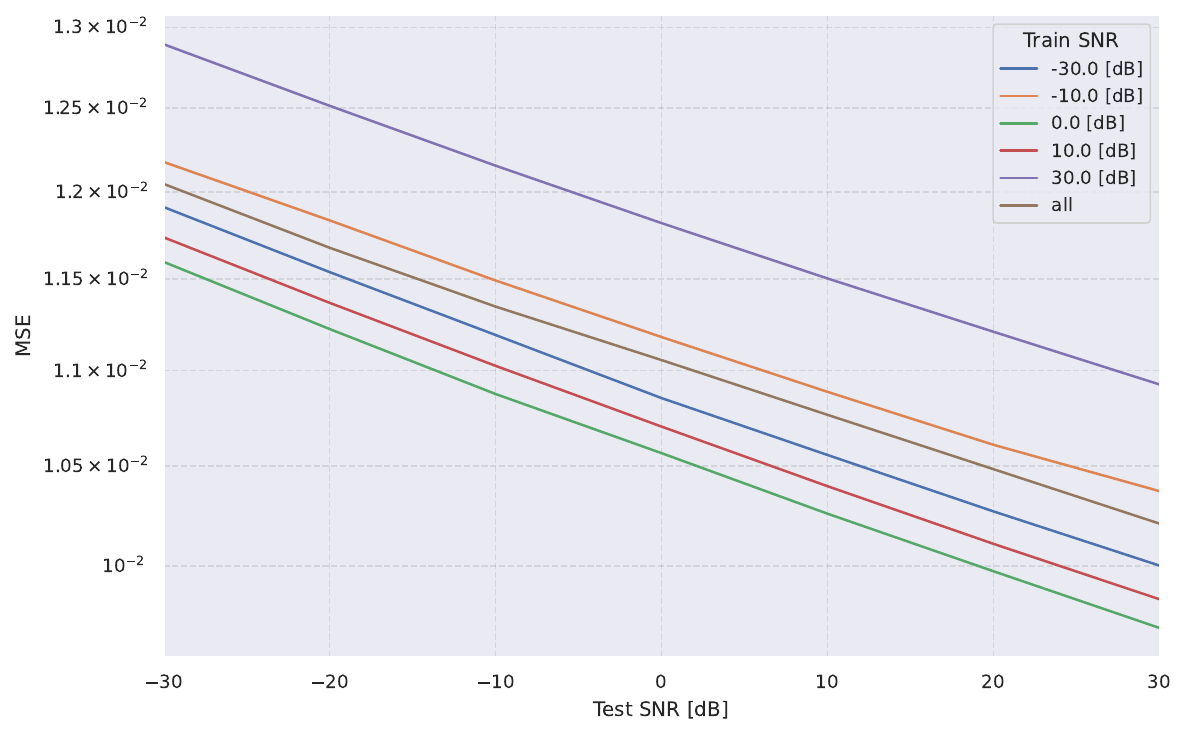}
         \caption{MSA, $v_{\textrm{train}} = 30$, $v_{\textrm{test}} = 0$}
         \label{fig:uma-slot-msa-trainV-30-testV-0}
     \end{subfigure}
     \begin{subfigure}[b]{0.49\textwidth}
         \centering
         \includegraphics[scale=0.37]{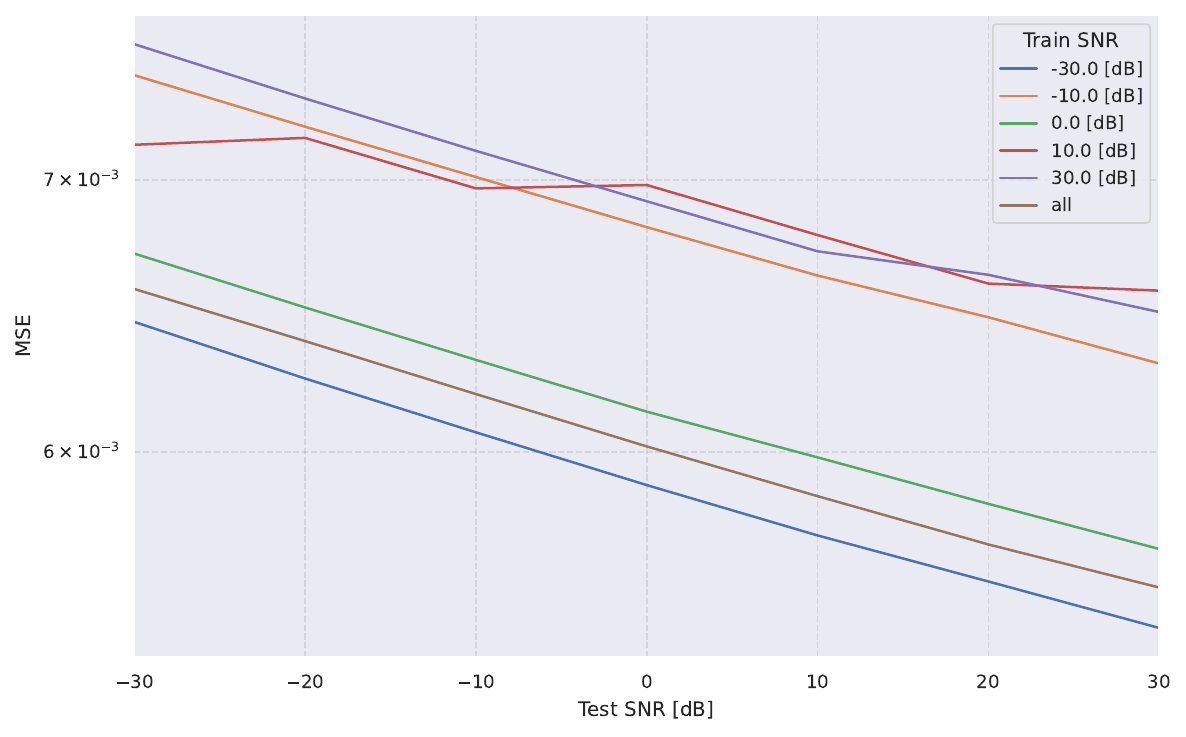}
         \caption{SSM, $v_{\textrm{train}} = 30$, $v_{\textrm{test}} = 0$}
         \label{fig:uma-slot-ssm-trainV-30-testV-0}
     \end{subfigure}
     \begin{subfigure}[b]{0.49\textwidth}
         \centering
         \vspace{0.2cm}
         \includegraphics[scale=0.37]{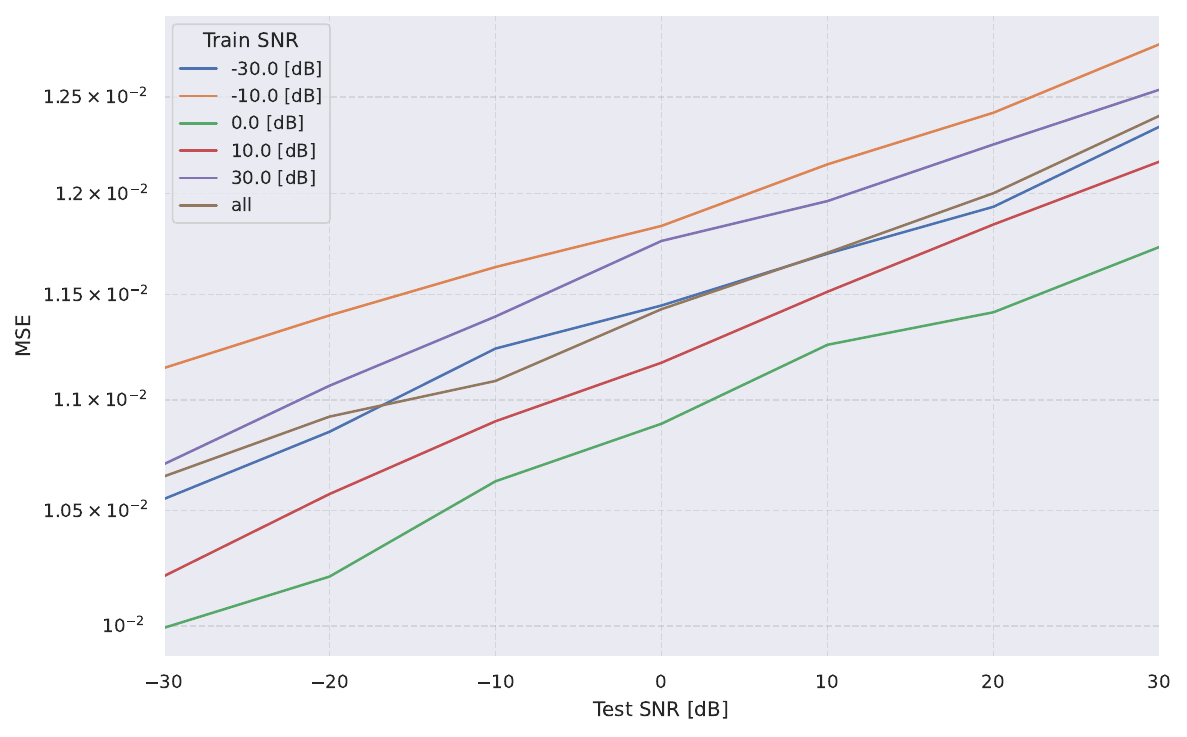}
         \caption{MSA, $v_{\textrm{train}} =0$, $v_{\textrm{test}} = 30$}
         \label{fig:uma-slot-msa-trainV-0-testV-30}
     \end{subfigure}
     \begin{subfigure}[b]{0.49\textwidth}
         \centering
         \includegraphics[scale=0.37]{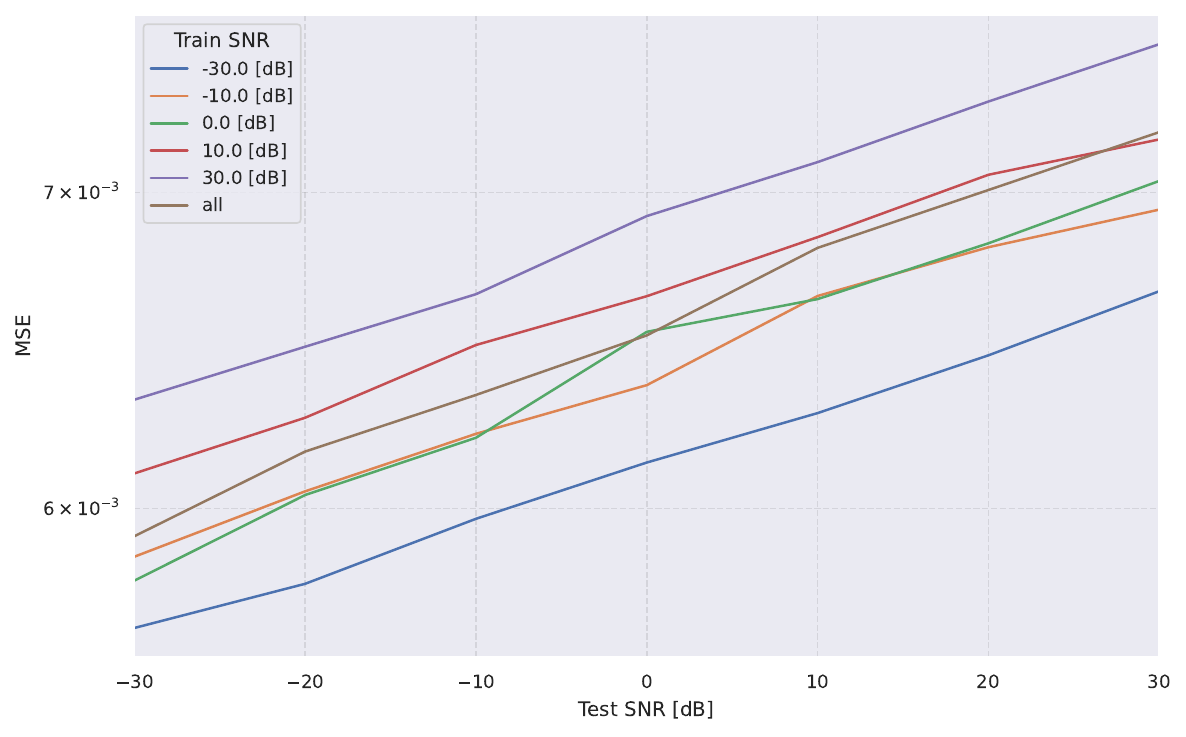}
         \caption{SSM, $v_{\textrm{train}} =0$, $v_{\textrm{test}} = 30$}
         \label{fig:uma-slot-ssm-trainV-0-testV-30}
     \end{subfigure}
    \caption{SISO MSE of next-slot OFDM-CSI prediction task vs. test SNRs at $f_c=5$ GHz for multiple MSA layers in (a) and (c) and the SSM layers in (b) and (d) when each is trained with the UMa channel at different SNR values \textit{with a distribution shift} in the UE speed (i.e., $v_{\textrm{train}} \neq v_{\textrm{test}}$).}
    \label{fig:uma-slot-OOD-0-30}
    %\vspace{-0.15cm}
\end{figure}

\end{appendices}

\end{document}